\documentclass[a4paper,fleqn,usenatbib]{mnras}
\usepackage[T1]{fontenc}
\usepackage{ae,aecompl}
\usepackage{graphicx}
\usepackage{amssymb}
\usepackage{amsmath}
\usepackage{bm}
\usepackage{mathptmx}
\usepackage{epstopdf}
\usepackage{booktabs}
\usepackage{float}
\usepackage{subfig}
\usepackage{epsfig,color,ulem}
\usepackage{tabularx}

\newcommand{\kms}{\,{\rm km s$^{-1}$}}

\mathchardef\mhyphen="2D

\newcommand{\civ}{C\,{\sc iv}}

\def\nh{\ifmmode n_\mathrm{\scriptscriptstyle H} \else $n_\mathrm{\scriptscriptstyle H}$\fi}
\def\ne{\ifmmode n_\mathrm{\scriptstyle e} \else $n_\mathrm{\scriptstyle e}$\fi}
\def\Qh{\ifmmode Q_\mathrm{\scriptstyle H} \else $Q_\mathrm{\scriptstyle H}$\fi}
\def\Uh{\ifmmode U_\mathrm{\scriptstyle H} \else $U_\mathrm{\scriptstyle H}$\fi}
\def\Nh{\ifmmode N_\mathrm{\scriptstyle H} \else $N_\mathrm{\scriptstyle H}$\fi}
\def\Uhhp{\ifmmode U_\mathrm{\scriptstyle H,HP} \else $U_\mathrm{\scriptstyle H,HP}$\fi}
\def\Nhhp{\ifmmode N_\mathrm{\scriptstyle H,HP} \else $N_\mathrm{\scriptstyle H,HP}$\fi}
\def\Uhvhp{\ifmmode U_\mathrm{\scriptstyle H,VHP} \else $U_\mathrm{\scriptstyle H,VHP}$\fi}
\def\Nhvhp{\ifmmode N_\mathrm{\scriptstyle H,VHP} \else $N_\mathrm{\scriptstyle H,VHP}$\fi}
\def\Nion{\ifmmode N_\mathrm{\scriptstyle ion} \else $N_\mathrm{\scriptstyle ion}$\fi}

\def\Zsun{\ifmmode {\rm Z}_{\odot} \else Z$_{\odot}$\fi}
\def\Msun{\ifmmode {\rm M}_{\odot} \else M$_{\odot}$\fi}
\def\kms{\ifmmode {\rm km~s}^{-1} \else km~s$^{-1}$\fi}
\def\Lya{\ifmmode {\rm Ly}\alpha \else Ly$\alpha$\fi}
\def\Lyb{\ifmmode {\rm Ly}\beta \else Ly$\beta$\fi}
\def\Lyg{\ifmmode {\rm Ly}\gamma \else Ly$\gamma$\fi}
\def\Lyd{\ifmmode {\rm Ly}\delta \else Ly$\delta$\fi}
\def\neaod{\ifmmode n_\mathrm{\scriptscriptstyle AOD} \else $n_\mathrm{\scriptscriptstyle AOD}$\fi}
\def\necrit{\ifmmode n_\mathrm{\scriptstyle cr} \else $n_\mathrm{\scriptstyle cr}$\fi}
\def\ncr{\ifmmode n_\mathrm{\scriptstyle cr} \else $n_\mathrm{\scriptstyle cr}$\fi}
\def\nepi{\ifmmode n_\mathrm{\scriptscriptstyle PI} \else $n_\mathrm{\scriptscriptstyle PI}$\fi}
\def\gtorder{\mathrel{\raise.3ex\hbox{$>$}\mkern-14mu\lower0.6ex\hbox{$\sim$}}}
\def\ltorder{\mathrel{\raise.3ex\hbox{$<$}\mkern-14mu\lower0.6ex\hbox{$\sim$}}}

\def\gsim{ \lower .75ex \hbox{$\sim$} \llap{\raise .27ex \hbox{$>$}} }
\def\lsim{ \lower .75ex\hbox{$\sim$} \llap{\raise .27ex \hbox{$<$}} }
\def\app#1#2{%
	\mathrel{%
		\setbox0=\hbox{$#1\sim$}%
		\setbox2=\hbox{%
			\rlap{\hbox{$#1\propto$}}%
			\lower1.1\ht0\box0%
		}%
		\raise0.25\ht2\box2%
	}%
}

\title[]{BALQSO Spectra Explained by Shock Disruption of Galactic Clouds}
\author[Meir Zeilig-Hess et al.]{
	Meir Zeilig-Hess$^1$\thanks{Email: meirzh@mail.tau.ac.il}, Amir Levinson$^1$, Xinfeng Xu$^2$,
	Nahum Arav$^2$
	\\
	$1$ The Raymond and Beverly Sackler School of Physics and Astronomy, Tel Aviv University, Tel Aviv 69978, Israel\\
	$2$ Department of Physics, Virginia Tech, Blacksburg, VA 24061, USA\\
}

\pubyear{2019}

\begin{document}
	\label{firstpage}
	\pagerange{\pageref{firstpage}--\pageref{lastpage}}
	\maketitle	
	
	\begin{abstract} 
Blue-shifted Broad Absorption Lines (BALs) detected in quasar's spectra are indicative of AGN outflows. We show, using 2D hydrodynamical simulations,  that disruption of interstellar clouds by a fast AGN wind can lead to formation of cold, dense high speed blobs that give rise to broad absorption features in the transmission spectrum of the AGN continuum source. For a wind velocity of $0.1 c$ and sufficiently high cloud density ($n_c\gsim10^4$ cm$^{-3}$, depending on size), absorption troughs with velocities up to about $3000$ \kms can be produced.  For slower winds and/or lower cloud density the anticipated velocity of the absorbing clouds should be smaller.

	\end{abstract}
	
\section{Introduction}

Quasars  show ubiquitous outflows \cite[$\simeq$ 20--50\% of all AGN:
e.g.][]{Hewett03,Dai08}, where blueshifted absorption lines are
attributed to sub-relativistic ($\sim10^3-10^4$ \kms) mass ejection.
These outflows are  a prime candidate for producing various AGN
feedback processes:
curtailing the growth of the host galaxy, explaining the relationship
between the masses of the central black hole and the galaxy's bulge,
and ICM and IGM chemical enrichment
\cite[e.g.][]{Ostriker10,Faucher-Giguere12,Zubovas14,Thompson15,Yuan18,zeilig2019}.

Most dynamical models of BAL winds assume that the outflow originates
from the AGN accretion disk
at $R\sim0.01$ pc (about $\sim10^{16}$ cm) from the central source
\citep[e.g.,][]{Murray95,Proga00,Contopoulos17}.  These outflows
typically reach 90\% of their terminal velocity within 2-4 times the
starting $R$.  Observationally, BAL troughs have widths of thousands
of \kms, which cannot be explained by just the thermal width.
Therefore, the above models must produce the troughs in the
acceleration phase where $R\sim0.1$~pc \citep[e.g.,][]{Murray95}.
However, over the
past decade, analysis of more than 20 individual outflows measured
$R\sim10-1000$ pc
\citep{Moe09,Dunn10a,Bautista10,Aoki11,Borguet13,Chamberlain15,Chamberlain15b,Xu18a,Miller18}.
 Moreover,  two recent  surveys demonstrated that half of the typical
BAL outflows are situated at $R>100$ pc \citep{Arav18,Xu18b}.

Since accretion disk wind models are inadequate for producing observed
BAL troughs at $R\sim10-1000$ pc, we turn our attention to models that
produce the troughs at the distance where they are observed.
The pioneering analytical work on such models was conducted by
\citet{Faucher-Giguere12} (henceforth FG12), where the BALs are
''formed in situ in radiative shocks produced when a
quasar blast wave impacts a moderately dense interstellar clump along
the line of sight."
However, the complicated spatial and time-dependent behavior of such
interactions limits the usefulness of the analytical approach in deriving
the physical characteristics of such outflows.

In this paper we analyze the process of cloud disruption by a wind, and the subsequent evolution and properties of the cloud
debris, using 2D hydrodynamical simulations.   The output of the simulation is then used to produce synthetic spectra along different
lines of sight that mimic observed spectra.
In difference from FG12, we consider the interaction of a AGN wind with highly dense ISM clouds.    
FG12 envisaged that the cloud accelerates to its terminal velocity (roughly the wind velocity) before
being broken into cold, dense fragments (cloudlets) that absorb the quasar radiation.    However, our simulations indicate that under these conditions
the cloudlets do not have time to cool substantially.    Moreover, the mean number of absorbing cloudlets anticipated to be seen along a typical
line of sight is too small to account for the broad absorption features detected in BAL spectra. 
As will be shown below, these problems are alleviated when the cloud is sufficiently dense, since (i) the cooling time of the shocked cloud
is much shorter, and (ii) the mean number of cloudlets is vastly larger. 
While, as mentioned above, a moderately dense cloud is disrupted only after acquiring high velocity, a
highly dense cloud is dispersed well before being accelerated significantly.   
We find that  the cloud is nearly completely destroyed by the time the radiative shock has crossed it, 
and that its debris are accelerated  by the wind to high velocities over a similar timescale.   
Our analysis indicates that for sufficiently dense clouds ($n_c >10^4$ cm$^{-3}$) this process can produce absorption 
troughs in the velocity range between 1000 and 3000 \kms, but not much higher.

\section{ BAL outflow Model}

\begin{figure}
		\centering
		\includegraphics[width=8cm]{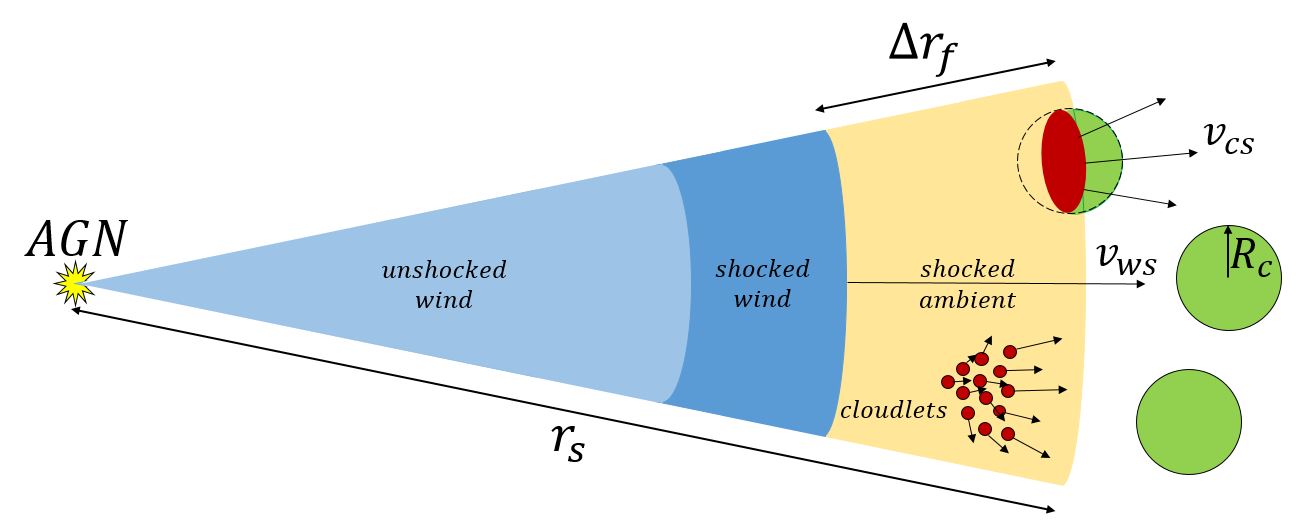}
		\caption[]{Schematic illustration of the wind-cloud model:  An AGN wind encounters a region in the galaxy containing
		overdense ISM clouds (the green circles).  Upon collision with a cloud, the latter is shocked and subsequently 
		shredded, by Kelvin-Helmholtz and Rayleigh -Taylor instabilities, into small fragments (indicated by the small red circles) that cool radiatively 
		over a short time.  These cloudlets are dragged 
		by the engulfing wind and accelerate to velocities that constitute a fraction of the original wind velocity.  
		The cold cloudlets absorb the quasar radiation, giving rise to  broad absorption troughs in the observed spectrum. The three zones
		indicated in the wind correspond to the layer of shocked ambient gas (of width $\Delta r_f$), the shocked wind and the unshocked wind. 
		}
		\label{fig:sketch}
	\end{figure}

The model considered below posits that a substantial sub-class of BAL outflows are composed of debris of
interstellar clouds that have been crushed by a fast AGN wind over a range of distances from the source
(parsecs to kiloparsecs) that vary from object to object (see Fig. \ref{fig:sketch} for illustration).   The exact nature of the wind's origin is
not specified and is not important for our analysis; it is merely assumed  that 
the wind propagates from its injection point, near the central engine, to 
a relatively large distance where it reaches a region in which dense molecular clouds are  abundant
and existing in a pressure equilibrium with their surroundings. 
Once the wind (more precisely, the shocked wind bubble) encounters a cloud, 
a strong shock wave is generated inside the cloud.  The shock
sweeps the cloud over a crossing time $t_{cc}$, which will
be calculated below, whereupon the cloud is broken into fragments (cloudlets).  
Each of these cloudlets is dragged and accelerated by the ram pressure of the wind to an
asymptotic velocity comparable to that of its surrounding bulk flow, over an acceleration time
$t_{drag}$, which will also be calculated below. During the dynamical
evolution just depicted, the dense cloudlets cool radiatively, and at the same time are heated by repeated 
shocks and compression waves.   This leads to a density-temperature relation down to 
a temperature at which dense enough cloudlets quickly cool to 
a temperatures between $10^4$ and $10^5K$.  At these temperatures there is a significant population 
of ions that produce the observed BAL (e.g., C IV, see section 4)

A particular BAL system is formed when intervening cloudlets along the line of sight absorb the quasar radiation.   
The spread of cloudlets encompasses a relatively small volume around the original location of the progenitor 
cloud and, therefore, the dynamics of the cloudlets is affected only by a small section of the wind, 
that which engulfed the original cloud.   To study the interaction of the wind and the cloudlets system it is thus
sufficient to specify the local wind quantities, rather than attempting to follow the global wind dynamics.  
This is the essence of the numerical model presented in section \ref{sec:simulations}.  
Nonetheless, to relate the BAL outflow model to global observables we start with
a brief description of the wind evolution in section \ref{sec:wind_dync}. 

The dynamics of cloud-wind interaction and cloudlets dragging is studied using 2D hydrodynamical simulations.  
Each simulation is run for long enough time to allow the cloudlets to accelerate to high velocities (a few thousands \kms).  The 
sizes, densities, velocities and spatial distribution of cloudlets obtained from the simulation are then used to calculate 
observed spectra along different lines of sight.  The details will be described in sections \ref{sec:simulations} and \ref{sec:spectra}.

	\label{lastpage}


\begin{figure*}
\includegraphics[width=8cm]{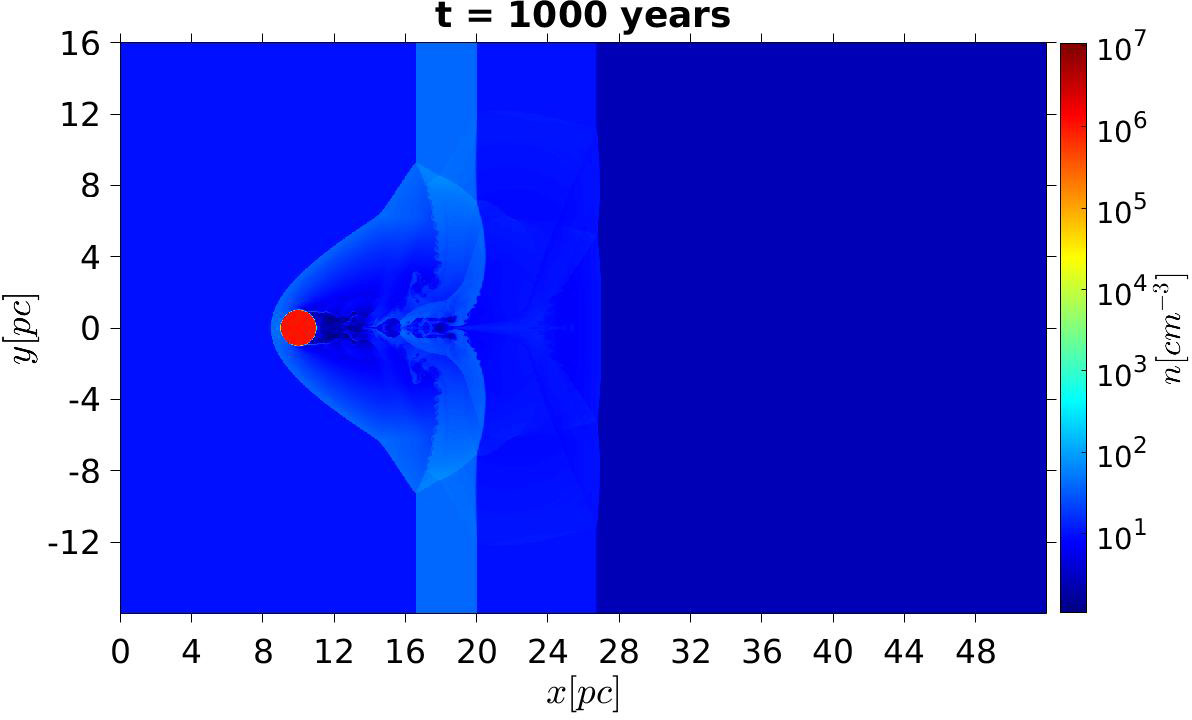} \includegraphics[width=8cm]{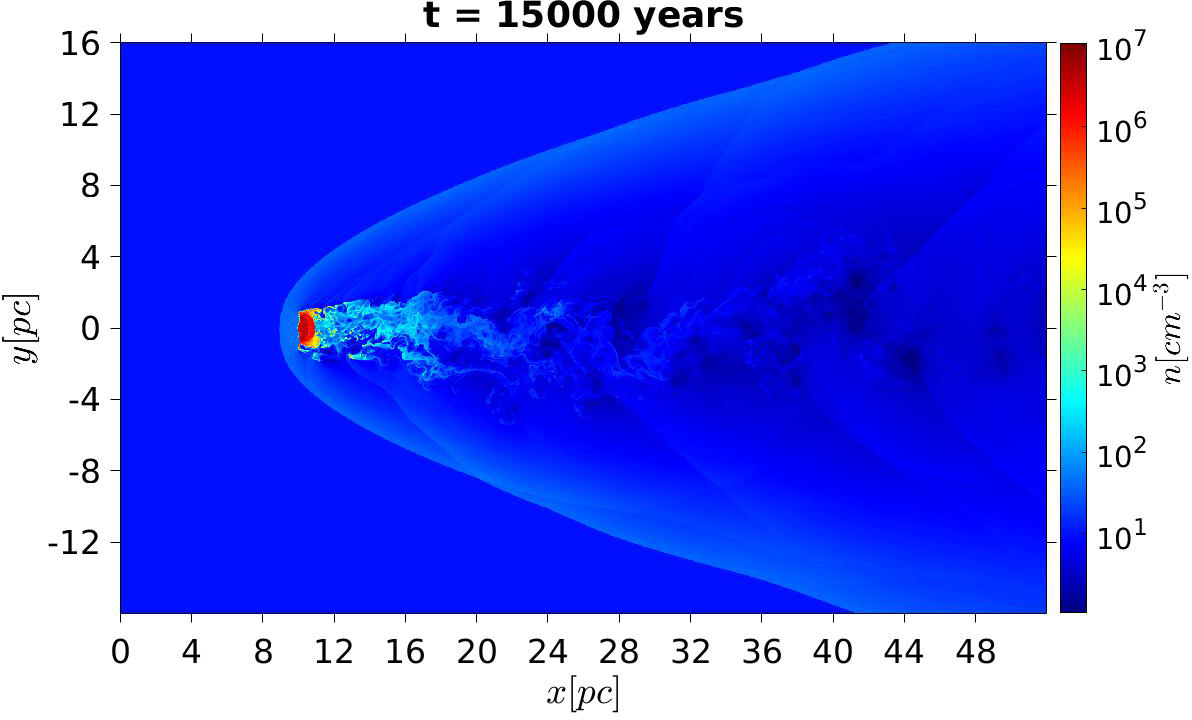}
\includegraphics[width=8cm]{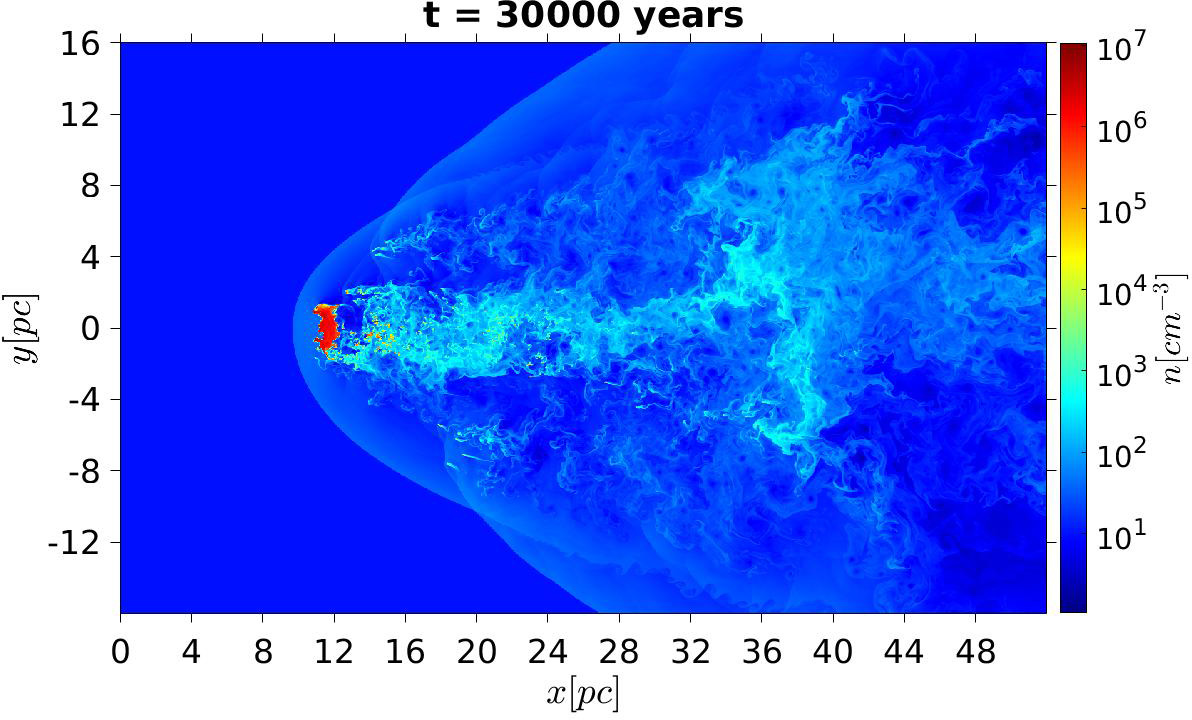} \includegraphics[width=8cm]{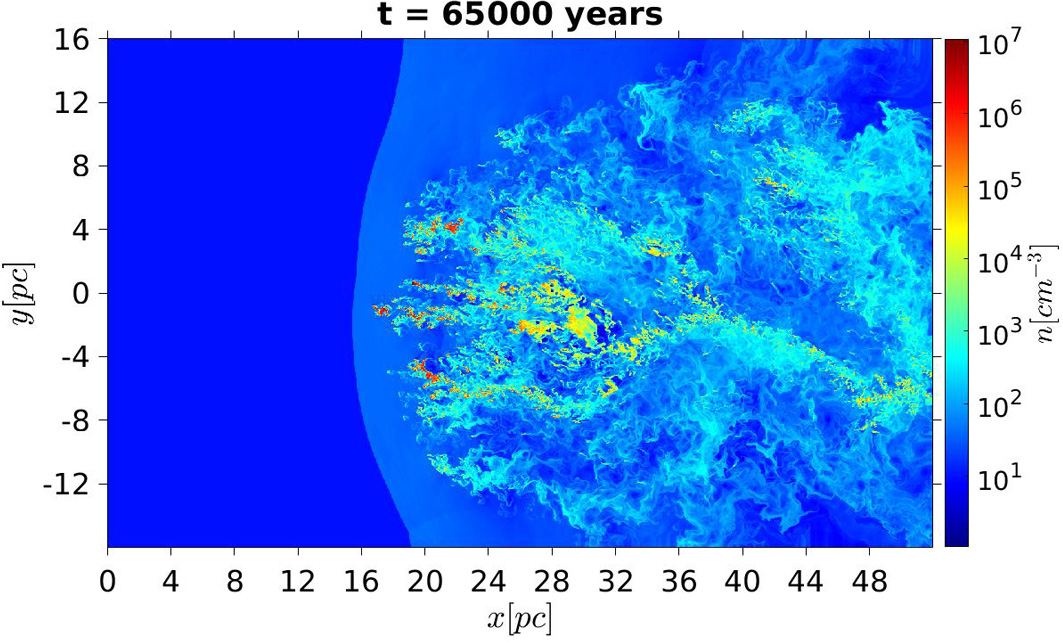}
\caption{Snapshots from the fiducial simulation showing density maps at four different times, as indicated.  Shock crossing 
is completed at $30000$ years.}
\label{fig:cloud_D}
\end{figure*}

\subsection{\label{sec:wind_dync} Wind dynamics}

Consider a conical AGN wind with a total kinetic power $L_{w}=10^{47}L_{w47}$ erg s$^{-1}$,
propagation velocity $v_{w}= \beta_w c$, and opening angle $\theta_w$.   The number density 
of the unshocked wind changes with distance $r =  r_{pc}$ pc from the wind source as 

\begin{equation}
n_{w}=\frac{L_{w}}{\pi\left(1-\cos\theta_w\right)m_{p}v_{w}^{3}r^{2}}=10^2 \frac{L_{w47}}{\left(1-\cos\theta_w\right)\beta_{w}^{3}} r_{pc}^{-2}\quad {\rm cm^{-3}},
\end{equation}
in the region where the wind remains approximately conical (no collimation). 
The wind is supposed to propagate
through an ambient medium with a power-law density profile of the form $n_{a}=n_{a0} r_{pc}^{-p}$,
with the exponent ranging from $p=0$ for a uniform background to $p=2$ for an isothermal spherical distribution,
and $n_{a0}$ is the density of the  ambient medium at a radius of 1 pc, henceforth measured in c.g.s units.
The interaction of the wind with the ambient gas leads to formation of a double-shock layer at the wind's head, 
within which the shocked wind and the shocked ambient gas are separated by a contact 
discontinuity (Fig \ref{fig:sketch}).  The shocked layer (more precisely the contact surface) propagates at a velocity that can be determined 
from local momentum balance to be $v_{ws}=v_{w}/\left(1+\sqrt{n_a/n_w}\right)$ \citep{zeilig2019}, and it is seen that 
significant deceleration of the wind's head commences at a radius $r_{dec}$, at which the ambient-to-wind density ratio exceeds unity. 
For the ambient density profile invoked above one finds:
\begin{equation}
r_{dec}\simeq\left[\frac{10^2~L_{w47}}{(1-\cos\theta_w)\beta_w^3n_{a0}}\right]^{1/(2-p)}\quad {\rm pc},
\end{equation}
and it is seen that for a wind velocity $\beta_w\lsim0.1$ and opening angle $\theta_w\simeq30^\circ$ substantial 
deceleration is anticipated at radii $r \gsim 1~{\rm kpc} (L_{w47}/n_{a0})^{1/2}$, adopting a uniform ambient 
density ($p=0$) for illustration.   The velocity of the wind's head (i.e., the double-shock layer) can be expressed in terms of $r_{dec}$ as 
\begin{equation}
v_{ws}=\frac{v_w}{1+(r/r_{dec})^{2-p}} \simeq ~ v_w(r/r_{dec})^{-(2-p)} ~ {\rm at }~ r>r_{dec}.
\label{eq:v_ws}
\end{equation}
It is worth noting that strong collimation of the wind is anticipated in the deceleration zone \cite[e.g.,][]{zeilig2019}, that will alter the density 
profile of the unshocked wind and, consequently, the head velocity $v_{ws}$.   Thus, at $r>r_{dec}$, $v_{ws}$ can be 
much larger than the value given by Eq. (\ref{eq:v_ws}).
Since observations of BAL outflows reveal cloudlets 
velocities up to about $0.1c$, it seems more likely to assume that the wind encounters the cloud
prior to decelerating, at $r \lsim r_{dec}$, although strong collimation may alter this inference. 

\subsection{Characteristic timescales}
A cloud encountered by the wind will first interact with the shocked ambient gas contained between the forward shock 
and the contact surface.    For a highly supersonic wind (i.e., a strong forward shock), the width of this layer is approximately 
$\Delta r_f\approx r_{s}/4$ when the forward shock reaches a radius $r_s$, and its crossing time is
\begin{equation}
t_f \approx \Delta r_f/v_{ws} \approx 10^3~{\rm yr}~ \left(\frac{r_s}{1~ {\rm kpc}}\right) \beta_{ws}^{-1}.
\label{eq:t_f}
\end{equation} 
This should be compared with the cloud crashing time which, as confirmed in section \ref{sec:simulations}  by numerical simulations, roughly
equals the crossing time of the shock generated inside the cloud by the cloud-wind collision.\footnote{ We find this result to hold 
even when the shock is very weak.}  
A strong shock is expected to form inside the cloud if the ram pressure of the shocked ambient flow, $m_pn_{as}v_{ws}^2$ (assuming
pure H composition for simplicity), largely exceeds the pressure of the unshocked cloud, $p_c=n_ckT_c$, where $n_c$ and $T_c$ 
are the initial density and temperature of the cloud. 
Expressed in terms  of the sound speed of the unshocked cloud, $c_s=\sqrt{\gamma kT_c/m_p}$, here $\gamma=5/3$ is the adiabatic index, and the
ratio $\kappa \equiv n_c/n_{a}$ between the cloud density $n_c$ and the (unshocked) ambient medium density $n_a = n_{as}/4$, 
the latter condition reads: $\kappa \ll 4\gamma v_{ws}^2/c_s^2$.
For a spherical cloud of radius $R_c$, the shock crossing time of the cloud is given by
\begin{equation}
t_{cc} = \frac{2R_{c}}{v_{cs}}\approx\left(\frac{R_{c}}{v_{ws}}\right)\sqrt{\kappa}\approx t_f\left(\frac{4 R_c}{r_s}\right)\sqrt{\kappa},
\label{eq:t_cc}
\end{equation}
noting that for a large density ratio ($\kappa>>1$)  the velocity of the shock inside the cloud is approximately 
$v_{cs}= v_{ws}\sqrt{n_{as}/n_c}=2v_{ws}/\sqrt{\kappa} > c_s$. 
Thus, as long as $\sqrt{\kappa}< r_s/4R_c$ the cloud will be swept by the shock before it reaches the contact surface. 
Otherwise, the disruption will continue in the wind itself (either the shocked or unshocked wind), whereby in the expression for $\kappa$,
$n_{a}$ must be  replaced by $n_{w}$ when the cloud is inside the shocked wind layer 
or by $4n_w$ when in the unshocked wind.  
Our numerical simulations show excellent agreement with Eq.(\ref{eq:t_cc}). 

Adopting for illustration $L_{w47}/n_{a0}=1$, $\beta_w=0.1$ and $\theta_w=30^\circ$ yields $r_{dec}= 1$ kpc and $t_f\approx 10^4$ yr at $r_{dec}$.
A cloud of radius $R_c=1$ pc will then be crashed at this distance within this time if its density satisfies $n_c \lsim 6\times10^4 n_{a0}$.

Another important timescale is the acceleration time of a dragged cloudlet, defined as the time it takes the cloudlet to reach the velocity of the shocked ambient layer.  To be precise, $t_{drag}=v_{ws}/a$, where $a=F/M_{ct}$ is the acceleration due to the drag force exerted on a cloudlet 
of mass $M_{ct}$ by the ram pressure of the shocked flow, $F=m_pn_{as}v_{ws}^2(\pi d^2)$.
For a spherical cloudlet of radius $d$ and density $n_{ct}$ the mass is given by $M_{ct}=m_pn_{ct}(4\pi d^3/3)$,
yielding
\begin{equation}
t_{drag} \approx \frac{4n_{ct} d}{3n_{as} v_{ws}} \approx t_f\left(\frac{16 d n_{ct}}{3r_s  n_{as}}\right).
\label{eq:t_drag}
\end{equation}
In practice the dragging time can be shorter by a factor of up to a few if the cloud is oblate. 
As an example, the acceleration time of a cloudlet having a density comparable to that of the progenitor cloud, $n_{ct}\sim n_c$, 
relative to the shock crossing time of the pre-crashed cloud is $t_{drag}/t_{cc}\sim (d/R_c)\sqrt{\kappa}$.
In \S \ref{sec:simulations} we find that the density distribution of the cloudlets span a wide range, owing
to large variations in the confining pressure in the region that contains the fragments of the crashed cloud.

\subsection{Cooling}
\label{sec:cooling}
For a (forward shock) velocity of $\beta_{ws}\lsim 0.1$ the temperature of the shocked ambient medium 
is $T_{as}\approx 10^{10} K (\beta_{ws}/0.1)^2$.  The dominant cooling process at such high temperatures
is free-free emission.   The corresponding cooling time is 
\begin{equation}
t_{cool}\approx 3\times 10^8~ {\rm yr}~ \left(\frac{T_{as}}{10^{10} K}\right)^{1/2}\frac{1}{n_{as}},
\end{equation}
much longer than the wind expansion time.   However, the dense cloud material cools over a much shorter time. 
Assuming the shocked cloud material to be in pressure balance with its surroundings, its
temperature is $T_{cs} = T_{as}(n_{as}/n_{cs})\approx T_{as}/\kappa$, and its free-free cooling time is shorter by 
a factor $\kappa^{3/2}$ than the cooling time of the shocked ambient gas,  adopting for illustration the jump conditions of a 
strong, non-radiative shock ($n_{cs}=4n_c$), for which $n_{as}/n_{cs}=4n_a/4n_c=1/\kappa$.
The above estimate holds at temperatures $T_{cs}>4\times10^7$ K.   At lower temperatures, $10^5 \lsim T \lsim 4\times10^7$, 
the cooling rate is considerably enhanced. For an ionized, optically thin plasma the cooling rate at $T=10^6$ K is larger by 
nearly two orders of magnitude than the free-free rate, depending on metallicity \citep[e.g.,][]{sutherland1993,gnat2012}.      To avoid unnecessary complications, we  adopt in our simulations a cooling function of the form 
\begin{equation}
\Lambda(T)= 1.4\times 10^{-27} \eta T^{1/2}~{\rm erg~s^{-1}~cm^{-3}},
\label{eq:cool_func}
\end{equation}
with $\eta>1$ being a free parameter ($\eta=10$ is adopted in the fiducial case study in \S \ref{sec:simulations}).  This function overestimates 
the cooling rate in the shocked ambient gas and underestimates the cooling rate of the cloud and cloudlets. 
However, even with this correction the cooling time of the shocked ambient gas is much longer than flow expansion time,
so it doesn't alter the dynamics.  
From Eqs. (\ref{eq:t_cc}) and (\ref{eq:cool_func}) with $\eta=10$ we find that the cooling time of the shocked cloud 
will be shorter than the shock crossing time $t_{cc}$ if the cloud-to-ambient density ratio satisfies   
$\kappa > 500 (R_c/1~{\rm pc})^{-1/2} (\beta_{ws}/0.1)n_{a}^{-1/2}$.   This condition is fulfilled in all cases considered below.   
In those cases the shock inside the cloud is radiative and the shocked cloud is 
compressed to a density of $n_{cs}\approx 10^7 n_{as}(\beta_{ws}/0.1)^2(T_{cs}/10^4~ K)^{-1}$.

\subsection{Cloudlets statistics} 
\label{sec:statistic}

A key question concerning the BAL model proposed here is how many cloudlets are expected to be detected along 
a given sightline on average.    A simple estimate can be made upon assuming that most of the mass of the 
original cloud is divided between identical cloudlets, each having a radius $d$ and density $n_{ct}$.  The total number
of cloudlets is then  $N_{ct} = (n_c/n_{ct})(R_c/d)^p$, where $p=2$ for a 2D system (as in our simulations below) and 
$p=3$ in 3D.
Now, let us choose $x$ to be the direction of the wind prior to 
collision with the cloud, viz., $\pmb{\beta}_{ws} =\beta_{ws}\hat{x}$, take the center of the cloud to be at  the origin, and
denote by $D$ the distance from the $x$-axis (along the perpendicular direction) within which the majority of 
cloudlets are located long after accelerating to
high velocities.  Then, the mean number of cloudlets anticipated along any line of sight 
within $D$ is: $N_{ct}(d/D)^{p-1}=(n_c/n_{ct})(R_c/D)^{p-1}(R_c/d)$.

Typically, $D/R_c\sim$ a few ($R_c/D$ ranges from about 2 to 5 in the simulations presented below, depending on $\kappa$). 
Thus, for a size ratio of $R_c/d =10^2-10^3$, between a few to a few tens cloudlets should be seen along any sightline that 
intersects the cloudlet cluster (that, is, within $D$) if $n_{ct}\simeq n_c$\footnote{In the model of \citet{Faucher-Giguere12}
$n_c/n_{ct}\simeq10^{-3}$, hence less than one cloudlet is expected to be seen along any line of sight on average}.   However, as will be shown in \S \ref{sec:simulations},
the density ratio $n_c/n_{ct}$ depends on the distribution of the confining pressure behind the pre-crashed cloud, which in turn depends 
on the wind velocity $\beta_{ws}$ and on the ratio $\kappa$, and can vary by nearly two orders of magnitude.  We shall get back
to this in the next section.


\section{Numerical Simulations}
\label{sec:simulations}
The calculations of the cloud-wind collision process have been performed using 2D 
hydrodynamical simulations.   A flow is injected from a planar boundary 
of a lateral extent large enough to contain the spread of cloud debris but small compared with the wind radius, and     
collides with a uniform heavy cloud initially embedded in a uniform, dilute ambient gas 
representing the galactic medium.   We denote the velocity and density of that flow by 
$\beta_{ws}$ and $n_{as}$, respectively, but note that this flow may represent either the shocked ambient flow, if cloud crashing and 
cloudlets acceleration occur in that region, or the wind itself if the shock crossing time of the cloud is considerably
longer than the wind expansion time $t_f$ given in Eq. (\ref{eq:t_f}).   
The planar approximation of the initial flow
is justified by the small ratio between the cloud radius to the putative cloud's
distance from the wind's source, as explained in the previous section.  Cooling of the 
gas is also included in the simulations in a manner specified below.
Each simulation has been run for long enough time 
to allow complete destruction of the cloud and subsequent acceleration of the 
cloudlets to high velocities. 

The output data of the simulation gives the spatial distribution of cloud fragments
at any given time, their velocities, densities and temperatures.   This data is used
to produce column density histograms in velocity space along different lines of sight, that
serve us later (section Eq. (\ref{eq:N_of_v}))  in creating synthetic spectra that can be compared
with observed BAL spectra.
It also enables us to get a grasp on the general dynamical and thermal evolution
of the cloud fragments, and to understand which of the following
processes happens faster in different stages of the system:
further fragmentation accompanied by shock heating versus cooling.

\begin{figure}
\includegraphics[width=8cm]{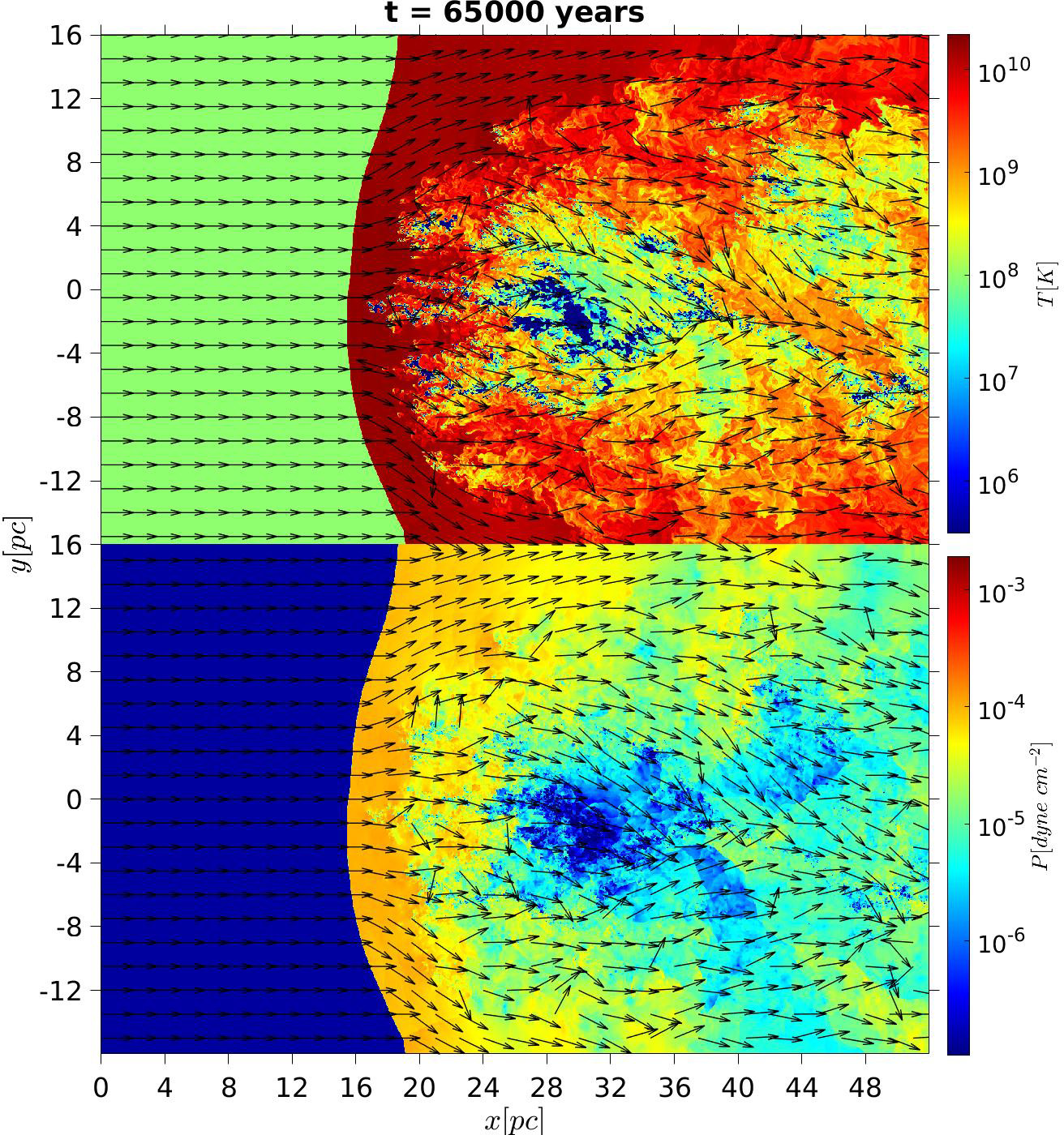}
\caption{Temperature (top) and pressure (bottom) distributions in the fiducial simulation at $t=65$ kyr.  The arrows indicate the velocity vectors of the flow.}
\label{fig:TemPres}
\end{figure}

\begin{figure*}[h]
\includegraphics[width=8cm]{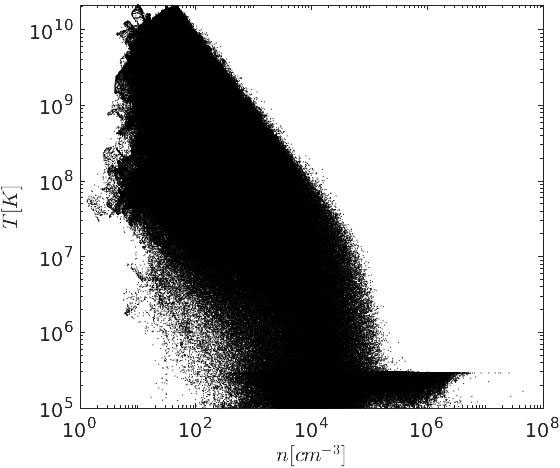}  \includegraphics[width=7.8cm]{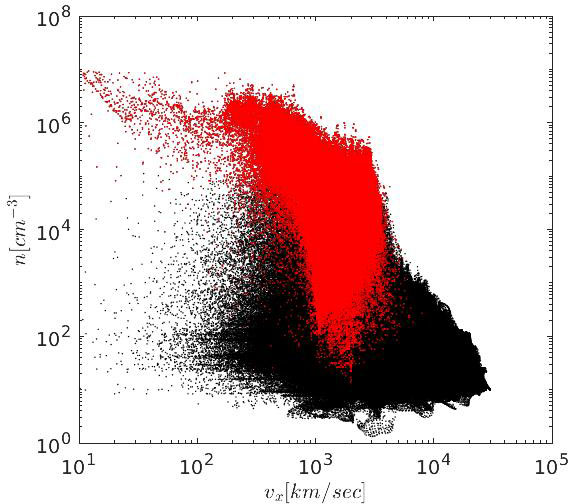}
\caption{Left: temperature-density phase diagram at $t=65$ kyr for all fluid elements in the fiducial simulation box. The relation $n\propto T^{-1}$ 
revealed in the diagram indicates pressure confinement.  The large scatter is due the pressure variation  inside the 
Mach cone, as seen in Fig \ref{fig:TemPres}.  The presence of points below the floor temperature is due to adiabatic expansion
of some dense cloud fragments.  Right: density-velocity diagram of the same matter.  The red points designate fluid elements at
temperature $T<10^6$ K.  As seen, the projected velocity of most fluid elements lies in the range $1000$ \kms $< v_x < 3000$ \kms. }
\label{fig:n-v}
\end{figure*}

\subsection{Setup}

We make use of the PLUTO code \citep{mignone2007} version 4.0
in 2D Cartesian coordinates in the non-relativistic HD module on an ideal gas with an adiabatic
index $\gamma=5/3$. 
We performed several experiments differing in the size of the simulation box, the implementation
of cooling and the ratio of densities, $\kappa$, between the cloud
and the wind. In each of these experiments, the ranges of the Cartesian
coordinates are of the form $0<x<x_{max}$ and $\left|y\right|<y_{max}$,
the unit length is equal to 1pc and the resolution along each Cartesian
direction is 0.01pc.  
The initial condition of the simulation is a static circular cloud
of uniform density $n_{c}$ with a radius $R_{c}=1$, surrounded by
a static ambient medium of uniform density $n_{a}$.   The cloud center is located
initially at $y=0$ and $x=x_c>0$, as close as possible to the wind source at $x=0$, but 
far enough to allow the bow shock created by the cloud-wind collision to remain in the 
simulation box at all times. 
The boundary condition at $x=0$ is a planar flow of uniform density
$n_{ws}$ moving at a speed of $v_{ws}$ along the $x$ direction
towards the cloud.   On the remaining boundaries we use open boundary conditions.
It is worth noting that in the absence of cooling the evolution of the system depends
only on the density ratio $\kappa=n_c/n_a$ and the velocity $\beta_{ws}$.  The inclusion
of cooling breaks down this scaling.   Moreover, the total mass of the original cloud determines the
absolute number of cloudlets produced following cloud crashing and, hence, the probability 
of observing BAL systems along different lines of sight around the original cloud location (see section \ref{sec:statistic} for discussion). 

Cooling is implemented by utilizing the Power\_Law switch with a bremsstrahlung
power law dependence of the free-free process, Eq. (\ref{eq:cool_func}).  We have made
runs with $\eta=0$ (no cooling), $\eta=1$ and $\eta=10$.  A floor temperature of $3\times10^{5}$
K is invoked to avoid excessive cooling.  As stated above, in reality the cooling rate shoots up
as the temperature drops below $10^6$ K, so that it can be safely assumed that cloudlets
that have reached the floor temperature will cool to $T\sim10^4$ K instantaneously.  

\begin{figure*}
\includegraphics[width=8cm]{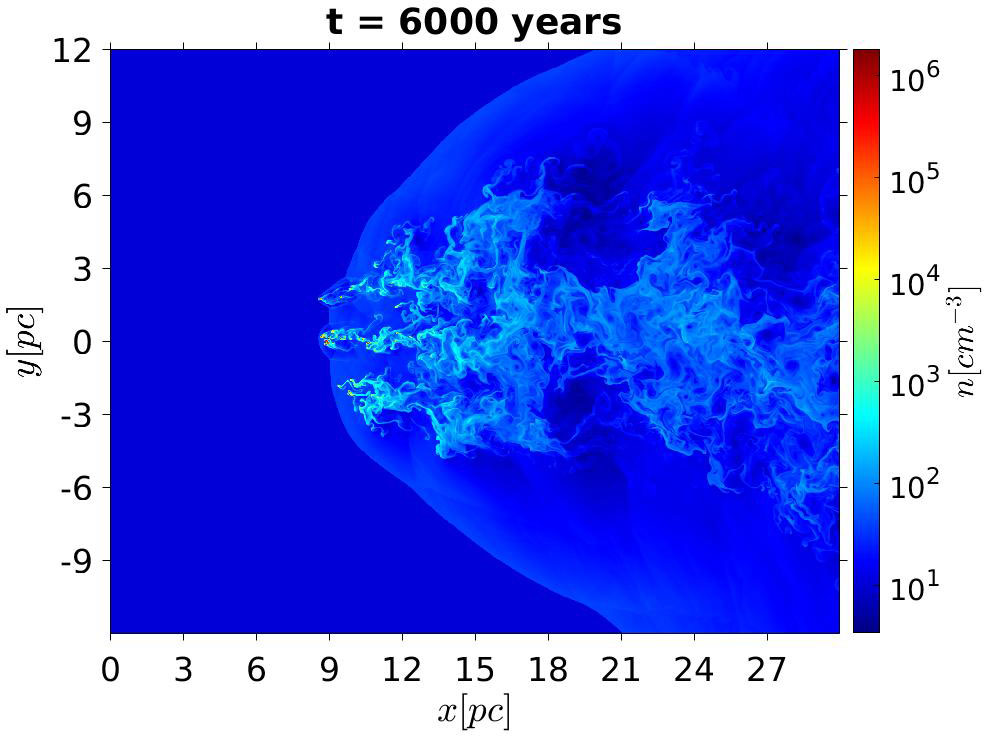}  \includegraphics[width=7.6cm]{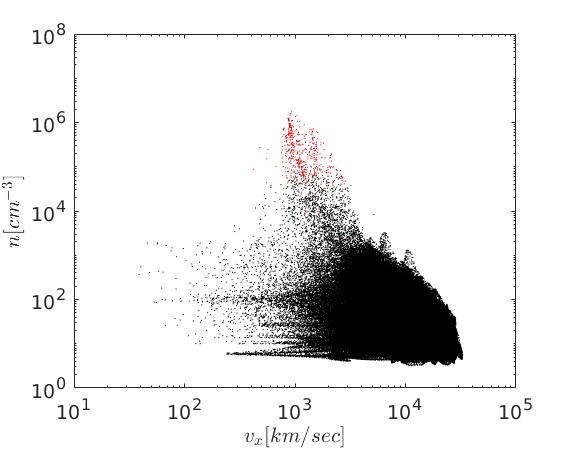}
\caption{Density map (left) and $n-v_x$ diagram (right) at time $t=6000$ yr for case B simulation. }
\label{fig:n-v_case2}
\end{figure*}

\subsection{Results}

We performed several experiments with different wind velocities, cloud densities, density ratio $\kappa$, and
cooling parameter $\eta$.  For our fiducial simulation we use $\beta_{ws}=0.1$, $\eta=10$, $\kappa=4\times10^5$ 
and cloud density $n_c=10^6$.    

Figure \ref{fig:cloud_D} shows a sequence of snapshots from the fiducial simulation.
The cloud crossing time in this run is 30,000 years, in good agreement with Eq. (\ref{eq:t_cc}). 
As seen, fragmentation of the cloud starts even before shock crossing is completed.  
A bow shock is formed at the moment of collision, and slightly expands as the cloud 
fragments (cloudlets) are scattered sideways, following complete destruction of the cloud.  
The bow shock decelerates the wind to a velocity smaller by up to a factor of a few than that of 
the injected wind $\beta_{ws}$, depending on location (see Fig \ref{fig:TemPres}).
This  limits the velocity to which the dragged  cloudlets can accelerate, as discussed further below. 
Figs \ref{fig:cloud_D}  (bottom right) and \ref{fig:TemPres} (top) reveal cold dense blobs embedded in the dilute background 
flow at time $t=65$ kyr, scattered over a transverse scale (perpendicular to the injected wind direction) 
about 5 times larger than the radius of the original cloud.  The 
velocities of these cold dense cloudlets span the range from zero to a few thousands km~s$^{-1}$ at this time. 
Figure \ref{fig:n-v} (right panel) depicts the density and the $x$ component of the velocity of each cell in the simulation box at time $t=65$ kyr.
Fluid elements having a temperature smaller than $10^6$ K are marked by a red color.   
The total number of these red points is about $3\times10^5$,
implying $\sim 300$ cold cloudlets on the average along any sight line that crosses the cluster, in the range $-5 <y/R_c < 5$.
Out of these about 72\%  have reached $v_x >10^3$ km s$^{-1}$, $16\%$  have reached $v_x>2000$ km s$^{-1}$ and 2\% 
$v_x>3000$ km s$^{-1}$.  A small number of cloudlets have reached even higher velocities,
up to $5000$ km s$^{-1}$.   In practice, the number of cold cloudlets is likely to be higher, by virtue of the stronger
cooling anticipated. 

The left panel in Fig. \ref{fig:n-v} shows the $n-T$ relation of all cells in the simulation box.   The slope of this relation, $n\propto T^{-1}$,
indicates that cloudlets are in pressure balance with their local environment.  However, there is a large spread in the density
of the filaments, that may look suspicious at first glance.   The reason for this is the large variation 
in the pressure behind the crashed cloud (see
bottom panel in Fig. \ref{fig:TemPres}), which is caused by the deflection of wind streamlines that engulf the dense cloud material. 
From Eq. (\ref{eq:t_drag}) we obtain $t_{drag}=3000 ~{\rm yr} (\beta_{ws}/0.1)^{-1}(n_{ct}/10^5~{\rm cm^{-3}})(d/0.01~{\rm pc})$ for a cloudlet of 
density $n_{ct}$ and size $d$ at the floor temperature.  Noting that the velocity of the background flow that drags the dense cloudlets is smaller than that of 
the injected wind by a factor of a few, we find that sufficiently small cloudlets  of density $n_{ct}\lsim 10^5$ can accelerate to the terminal 
velocity within about $30,000$ years, while the dragging time of considerably denser cloudlets is much longer, 
as indeed seen in Fig \ref{fig:n-v}.
The $n-T$ plot also reveal that all cloudlets with densities $n_{ct} \gsim10^5$ cm$^{-3}$ quickly cool to the floor temperature.    
Cloudlets of lower densities exhibit a wide range of temperatures, presumably due to heating by repeated shocks. 
In practice, we expect this sharp drop in temperature seen at $n\simeq 10^5$ cm$^{-3}$  in Fig \ref{fig:n-v} to 
occur at lower densities, since the cooling rate is likely to be substantially higher than that invoked in our simulations. 
The cooling of cloudlets to temperatures below the floor value, $T_f=3\times10^5$, K might be caused by adiabatic expansion
of dense cloudlets as they move into a low pressure zone. 

The qualitative behaviour of the system seen in the fiducial simulation is quite typical.  In all cases with $\beta_{ws}=0.1$ we find 
that most cold ($T_{ct}\le10^6$ K) dense cloudlets accelerate to velocities in the range $1000 - 3000$ km s$^{-1}$ 
right after shock crossing, at $t\sim 2t_{cc}$, with much slower acceleration at later times.   A very small fraction of the cloudlets accelerated 
to even higher velocities, up to $5000$ km s$^{-1}$ in the fiducial case.  For slower winds (injected velocity $\beta_{ws}=0.03$)
the terminal velocity of cloudlets was found to be somewhat smaller.   The total number of cloudlets depends also on the density
of the pre-crashed cloud through the ratio $n_c/n_{ct}$, as explained in section \ref{sec:statistic} above.

As another example (case B) we show in Fig. \ref{fig:n-v_case2} the density map and $n-v$  diagram at time $t=6000$ yr for a
simulation with the same wind parameters and cooling ($\beta_{ws}=0.1$, $\eta=10$), but a lower cloud density, $n_c=10^4$ cm$^{-3}$.
The corresponding density ratio is $\kappa=4\times 10^3$.  The cloud crashing time in this case is $t_{cc}=3000$ yr, as expected,
and it is seen that as in the fiducial simulation, cloudlets reach their terminal velocities at $2t_{cc}$. We have run the simulation up to 
a longer time but didn't observe a significant change in the velocity distribution of the cloudlets at times $t>6000$ yr.   The vertical 
spread (along the $y$ axis) is $D\simeq 2R_c$ (Fig. \ref{fig:n-v_case2}), smaller than that found in the fiducial simulation.  
This is anticipated given the smaller density ratio $\kappa$ in this run. 
The total number of dense cold cells is about 600, which translates to roughly $2$ cloudlets along sight lines in the range $-2 < y/R_c <2$,
compared with hundreds in the fiducial case.    
However, as explained in section \ref{sec:statistic}, this number is proportional to the initial cloud radius $R_c$, hence,
for a considerably larger pre-crashed cloud many more cloudlets should be seen in this case along sight lines that intersect the cloudlets cluster. 

\begin{figure}
\includegraphics[width=8cm]{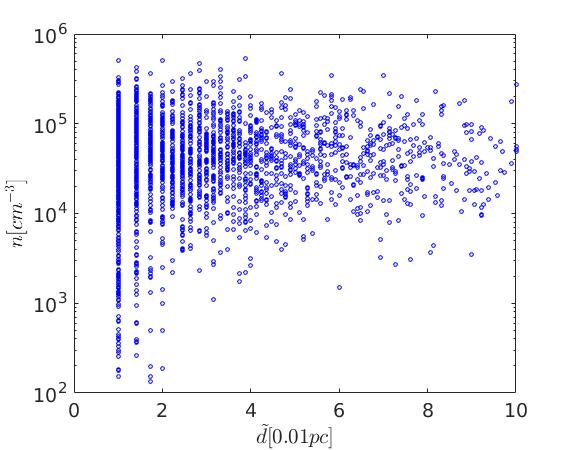}  
\caption{Mean size distributions of cloudlets at different densities.  The numbers on the horizontal axis (i.e., the values
of $\tilde{d}/(0.01$pc)) give the square root of the number of cells in the cloudlet.}
\label{fig:resolution}
\end{figure}

\subsection{Resolution}
The size of cloudlets in our simulations is limited by spatial resolution.  With our grid spacing the smallest cloudlets have size of $0.01$ pc.  
This means that further disruption of these cloudlets may be artificially suppressed.    To check the effect of numerical resolution on our 
results we computed the size distribution of cloudlets in the simulation box in the fiducial case at time $t=65$ kyr.   Fig \ref{fig:resolution} shows the
mean size distribution of cloudlets at different densities, where
the mean size $\tilde{d}$ is defined as $\tilde{d}=\sqrt{A_{ct}}$ in terms of the cloudlet area $A_{ct}$, so that $(\tilde{d}/0.01pc)^2$ is 
essentially the number of cells contained in the cloudlet. 
We find that only $18 \%$ of the cloudlets contain one cell, and that over $50 \%$ contain at least 10 cells.   This means that absorption 
occurs predominantly in resolved cloudlets\footnote{Note that a large cloudlet can contribute to absorption along several lines of sight.}.

\begin{figure}
\includegraphics[width=8cm]{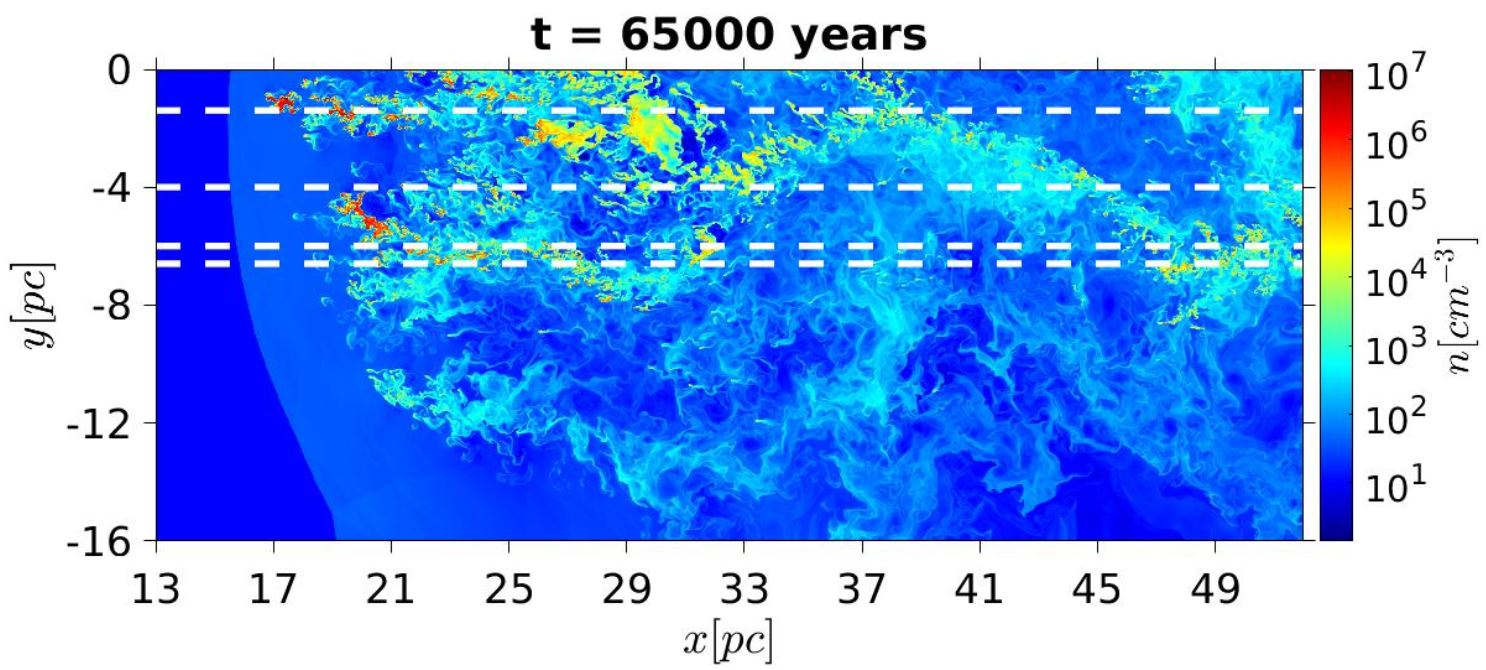}  
\caption{The white dashed lines delineate four different sight lines (y = -1.4, -4.0, -6.0, -6.6) in the fiducial simulation, 
along which sample absorption spectra are computed in section \ref{sec:spectra}.}
\label{fig:sightlines}
\end{figure}

\section{Synthetic Spectra}
\label{sec:spectra}

\begin{figure*}
\includegraphics[width=6.5cm]{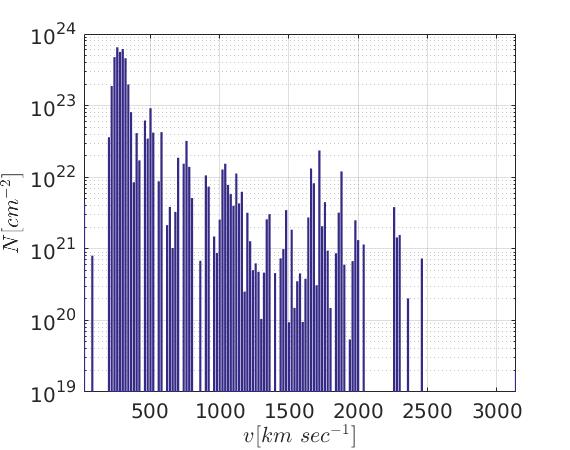}  \includegraphics[width=6cm]{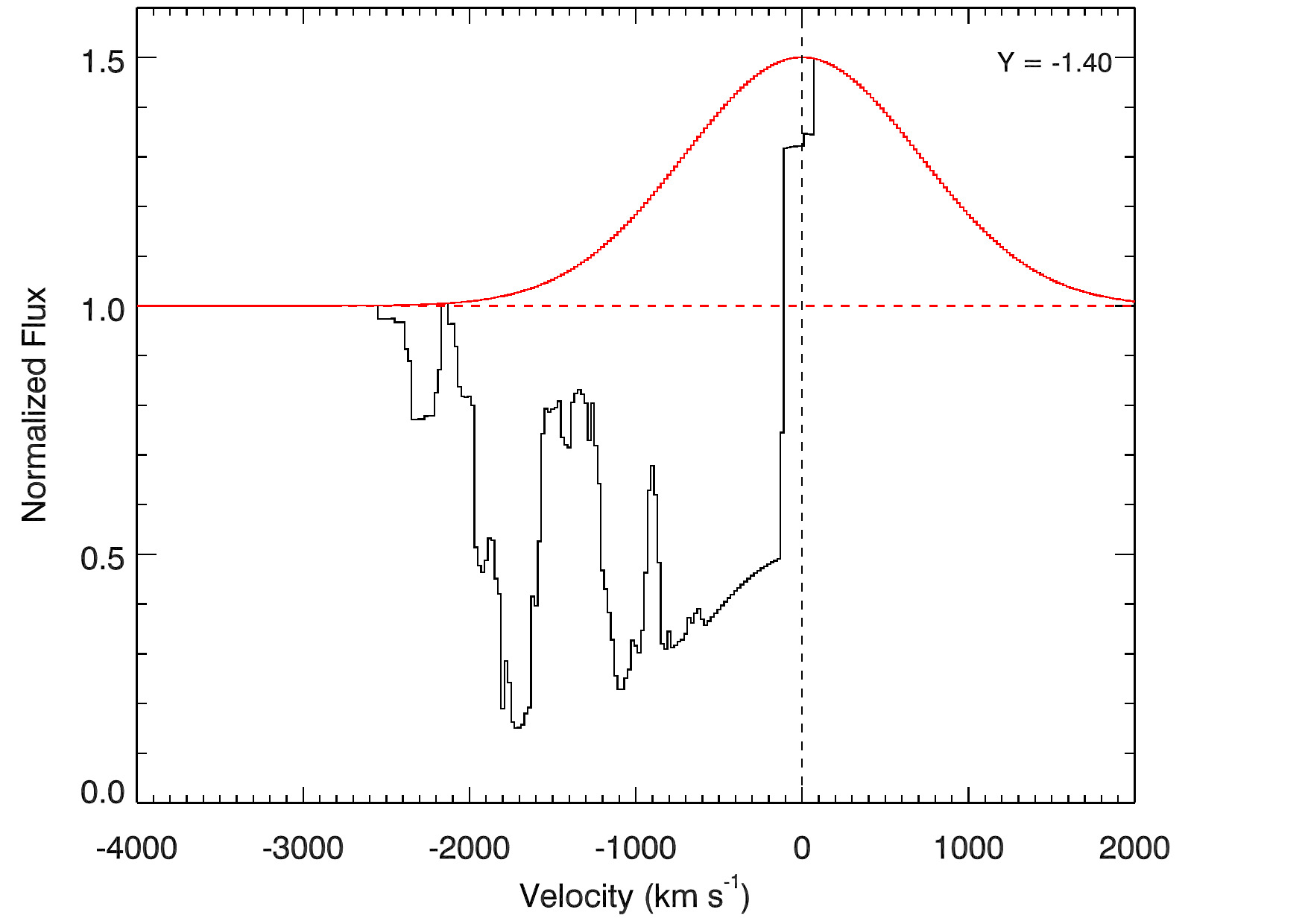}  \includegraphics[width=6.5cm]{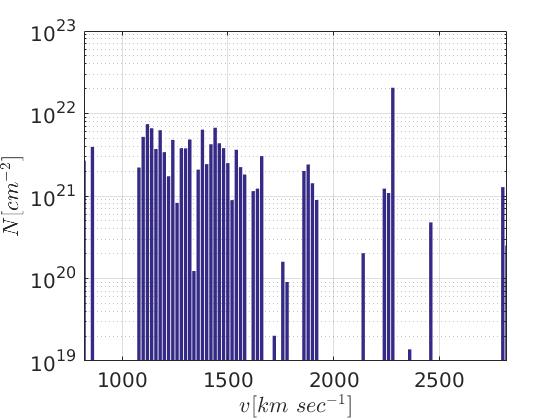}  \includegraphics[width=6cm]{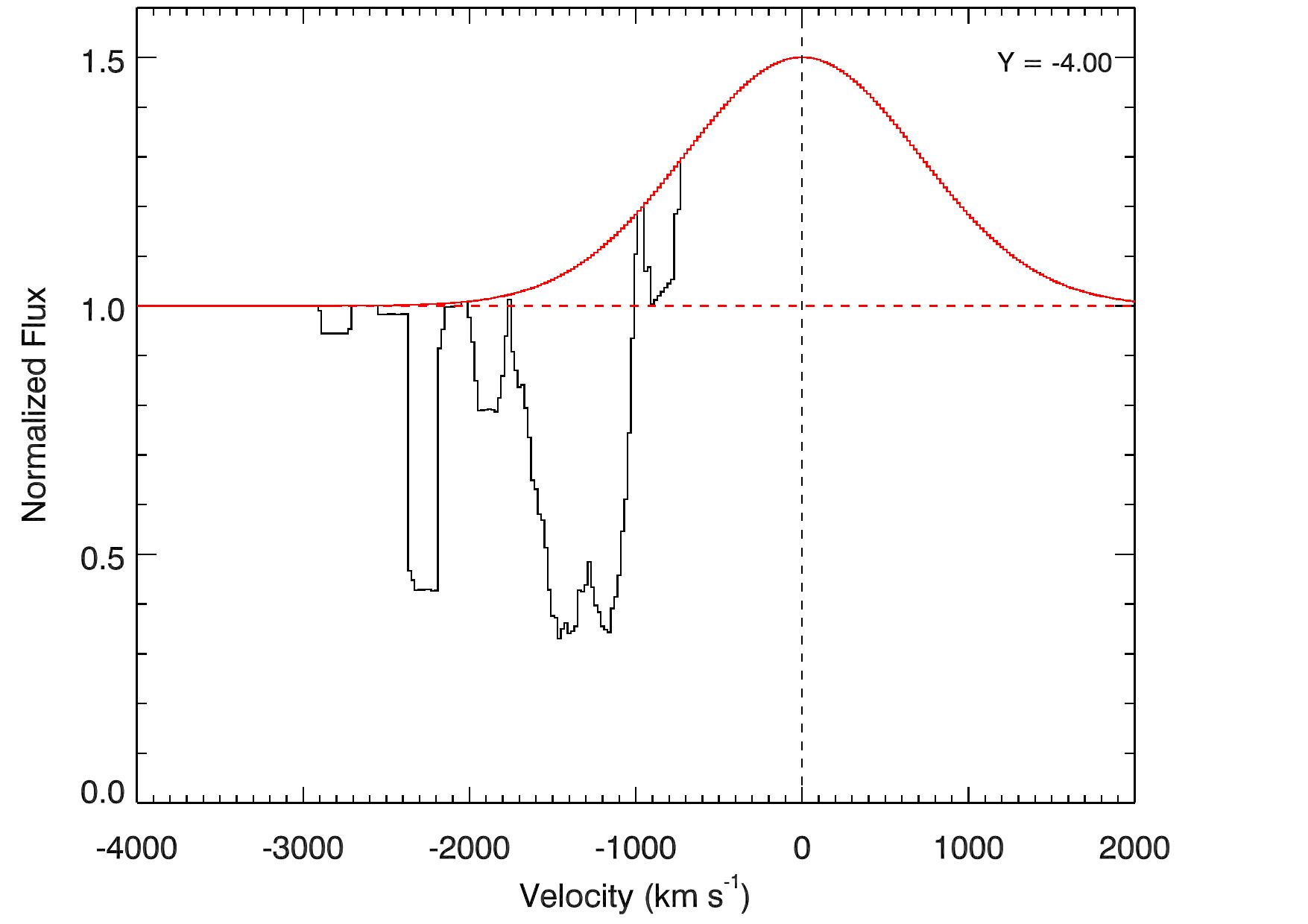}  
\includegraphics[width=6.5cm]{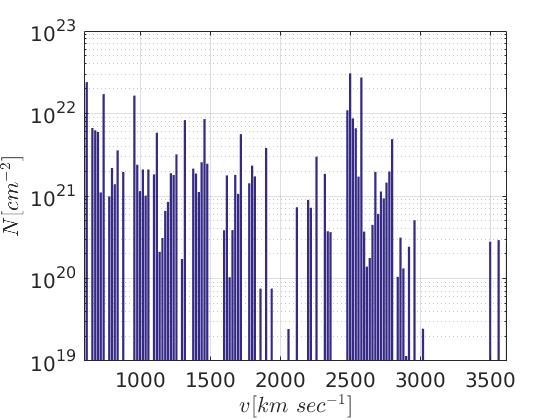}   \includegraphics[width=6cm]{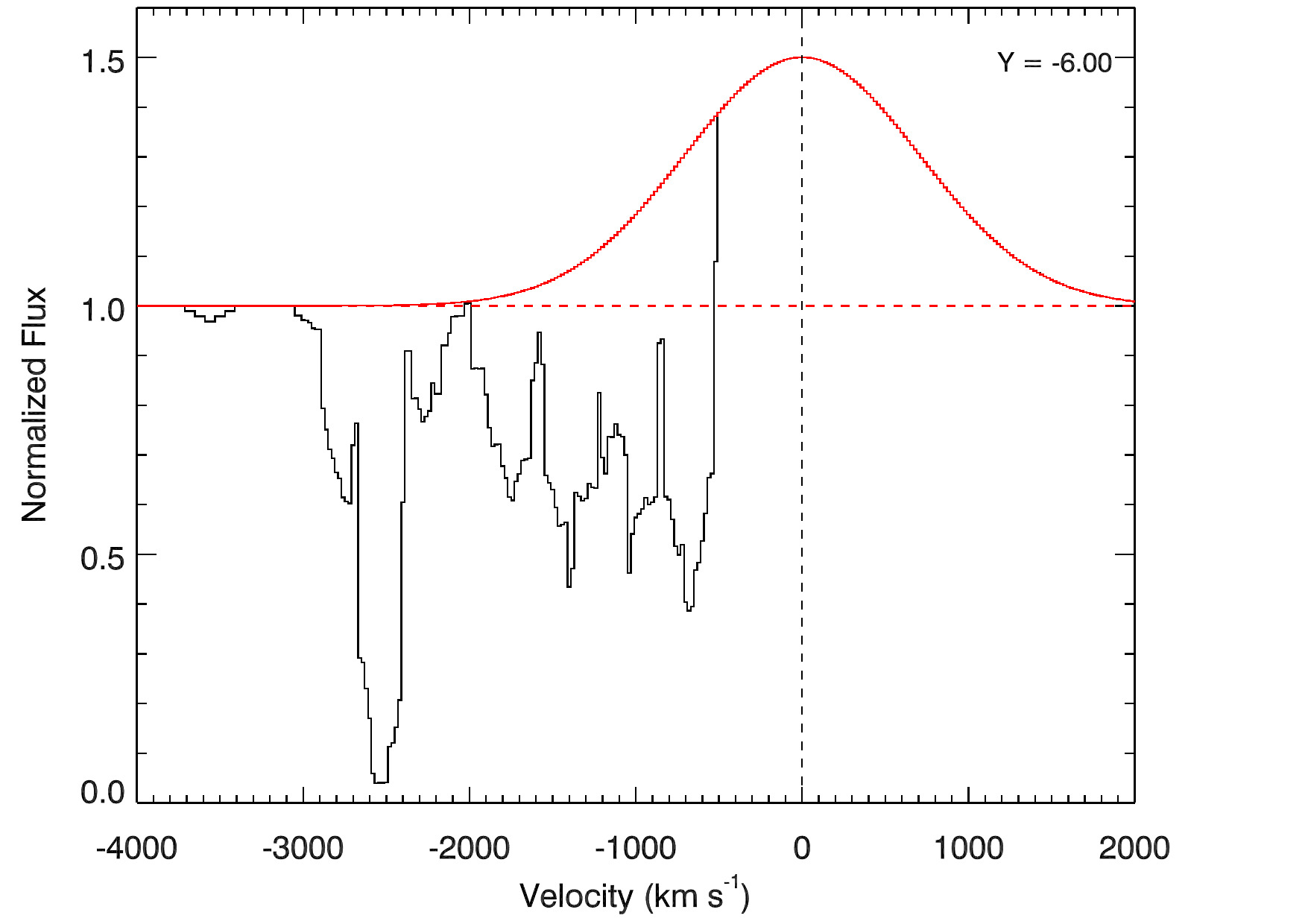}  \includegraphics[width=6.5cm]{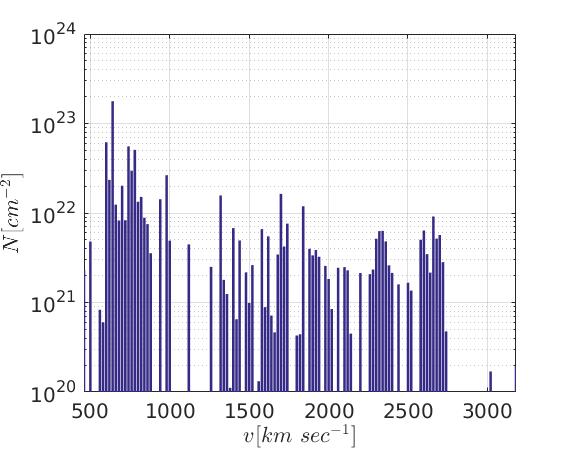}  \includegraphics[width=6cm]{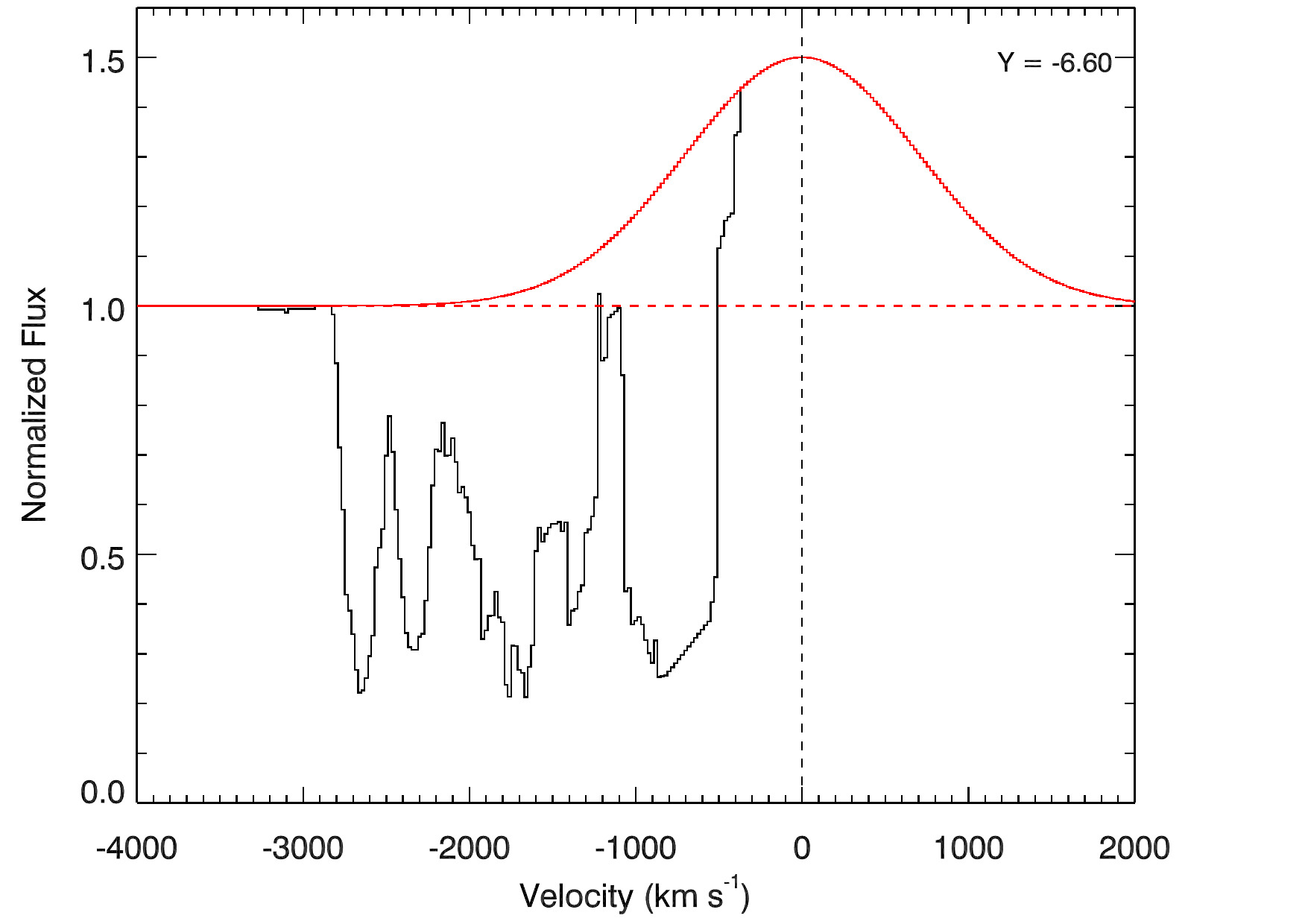}  
\caption{The left panels exhibit column densities of cold gas as a function of projected velocity $v_x$ along the lines of sight 
$y=-1.4, -4.0, -6.0, -6.6$ (from top to bottom).  
The width of each bar in the histogram corresponds to a velocity bin of $\Delta v_x=20$ \kms. 
The right panels show the corresponding transmission spectrum.}
\label{fig:N_v}
\end{figure*}

This section outlines the method employed to calculate absorption spectra along given lines of sight, using the simulation data.
Several representative examples of such spectra will be exhibited for illustration.
A more comprehensive analysis of absorption spectra and comparison 
to observations will be presented in a follow-up paper. 
For simplicity, we suppose that the wind is conical (as opposed to strongly collimated), so that absorption of the quasar
continuum radiation is seen along directions parallel to the wind streamlines.  Within our planar approximation, this means 
sight lines along the $x$-direction (each marked by a value of the $y$ -coordinate).  Examples are shown in Fig \ref{fig:sightlines}.
As explained in the preceding section, the planar approximation is justified by the fact that the observer is situated
at a distance much larger than the size of both the AGN source, the
original pre-shocked cloud and the system of spread-out cloudlets. 
We further assume that, in practice, fluid elements that have reached temperatures $T\lsim10^6$ K in the simulation 
will cool quickly to lower temperatures, $T\approx10^4$ K, by virtue of the much faster cooling  anticipated (see section \ref{sec:cooling} for discussion).  
When computing column densities that contribute to absorption we select only matter having $T<10^6$ K.

We shall focus here on the fiducial simulation, noting that the other cases lead to similar results, provided cloudlets 
are sufficiently abundant (see section \ref{sec:statistic} for a detailed discussion).  The
lines of sight indicated on the density map in Figure \ref{fig:sightlines} will be used to illustrate the method and 
produce sample spectra.  

The first step in the construction of synthetic BAL-spectra is the
calculation of the column-density as a function of the projected velocity along the observed direction, here $v_x$. 
Once the column-density is obtained, the optical depth 
for absorption at the corresponding  wavelength will be calculated, out
of which the normalized flux would be derived.
For any given sight line we integrate the number density
of cold matter ($T<10^6$ K) within a given velocity bin along the $x$-direction.  
In order to preserve the information of velocity dependence, we construct a 3D array of number-density
values, distributed in 2D configuration-space (i.e., in the x-y plane) and in 1D projected velocity
space, with a velocity bin size of $\Delta v_{x}$. 
Only after obtaining the phase-space distribution of number-densities,
we perform the horizontal integration to obtain the column density  of matter 
moving in the velocity range $(v_x, v_x+\Delta v_x)$, along a particular $y$-line:
\begin{equation}
N\left(y,v_{x},\Delta v_x\right)=\int_0^{x_{max}} dx \int_{v_x}^{v_x+\Delta v_x} \frac{d n\left(x,y,v_{x}\right)}{d v_x} dv_x.
\label{eq:N_of_v}
\end{equation}
In the integration we utilize a measure equal to the spatial resolution
limit of the simulation, $\Delta x=0.01\textrm{pc}$. The resulting
column-density (henceforth $N_{H}$) distributions in horizontal-velocity
space  with a bin size of $\Delta v_{x}=20$ \kms (chosen to match the resolution of the COS spectrograph on HST)
are shown in the left panels in Figure \ref{fig:N_v} for the four lines indicated in Fig. \ref{fig:sightlines} ($y= -1.4,-4.0,-6.0,-6.6 $).
It reveals substantial column densities up to $v_x \lsim~ 3000$ \kms. 

The sight-line analysis confirms that the typical values of $N_{H}$ and speeds do
not change much in the vertical direction, up until reaching large
enough transverse coordinates $|y| \gsim 6$, after which a significant
drop in the number of blobs is seen. This means that at the late stage
analyzed here, most of the cold-dense blobs are dispersed into a region
of size about 6 times that of the original cloud, as indeed seen by eye in the density map, Fig. \ref{fig:sightlines}.
The mean number of cloudlets identified along sight lines within the cloudlets cluster is consistent with the naive estimate 
in section \ref{sec:statistic}.


The most ubiquitous BAL in quasar spectra arises from \civ. Moreover, the defining criteria of a BAL depends on the \civ\ trough having velocity width larger than 2000 \kms\ at normalized residual intensity I=0.9 \citep{Weymann91}.  We therefore produce a synthetic spectrum that shows the \civ\ absorption that is derived from our fiducial simulation.
The observed \civ\ BAL arise from the doublet transitions at 1548.19\AA\ and  1550.77\AA. To highlight the simulation's results, we only use the stronger transition at 1548.19\AA.  In  reality, when the velocity width of the outflow is larger than the equivalent 500\kms\ velocity separation between the doublet components, the trough will be a blend of absorption from both transitions.

To produce a synthetic \civ\ BAL from our simulation we need to: \\
a) Determine $N_\text{CIV}{(y,v_x)}$ from the simulation's $N_H(y,v_x)$ (see Eq. (\ref{eq:N_of_v}) and accompanied discussion).\\
b) Derive the optical depth [$\tau_\text{CIV}{(y,v_x)}$] associated with $N_\text{CIV}{(y,v_x)}$.\\
c) Create a synthetic spectrum from the convolution of expected absorption
associated with $\tau_\text{CIV}{(y,v_x)}$ and the emission sources of the quasar

To Determine $N_\text{CIV}{(y,v_x)}$, we first need to take into account the abundance of carbon relative to hydrogen. For this purpose we use standard solar metallicity \cite[][]{GASS10}. Second, we
need to determine the fraction of \civ\ to all carbon atoms.
Ionization equilibrium in a quasar outflows is dominated by photoionization caused by the ionizing flux of the quasar \citep[e.g.,][]{Arav01}. For illustrative purposes we use the photoionization solution of the  outflow observed in
quasar HE 0238-1904 \citep[][]{Arav13}.  The solution was derived using
the spectral synthesis code Cloudy [version c17.00, \cite{Ferland17}].  As input we used the spectral energy distribution (SED) constructed from the observations of the quasar. The solution that best fits the data has log(\Uh) = 0.3 and log(\Nh) = 20.8 [log(cm$^{-3}$)] \citep{Arav13}, the photoionization solution yields N$_\text{CIV}$/\Nh\ $\simeq$ 9.0 $\times$ 10$^{-8}$. Therefore, we use:

\begin{equation}
\label{Eq:NCIV}
N_\text{CIV}(y, v_{x}) = 9.0 \times 10^{-8}\Nh(y, v_{x})
\end{equation}

The corresponding optical depth ($\tau$) of the  \civ\ absorption  can be expressed as \citep[see equation (8) in][]{Savage91}:

\begin{equation}
\label{Eq:tauCIV}
\tau_\text{CIV} (y, v_{x} ) = 2,654 \times 10^{-15} f\lambda \times N_\text{CIV}(y, v_{x})
\end{equation}

where for a \civ\ trough, $\lambda$ = 1548.19\AA\ and $f$ = 0.19 is the wavelength and the oscillator strength of the transition, respectively; and  
$N_\text{CIV}(y, v_{x})$ is in ions cm$^{-2}$(km s$^{-1})^{-1}$.

Quasars' ultraviolet emission is composed of a continuum source and a broad emission line (BEL) source, where the size of the latter is at least a hundred times larger than the former. Therefore, it is not surprising that in many cases the outflow covers the continuum source entirely, but only a negligible portion of the BEL emission \citep[e.g.,][]{Arav99}. Our synthetic spectra assume this empirical result: the absorber fully cover the continuum, but does not cover the BEL.

The flux of the simulated \civ\ absorber is therefore given by the absorbed continuum (normalized to 1):

\begin{equation}
\label{Eq:troughCIV}
I_\text{CIV} (y, v_{x} ) = e^{-\tau_\text{CIV}(y, v_{x} )}
\end{equation}
and an added unabsorbed BEL whose parameters are: velocity centroid at 0 km s$^{-1}$, $\sigma$ = 1000 km s$^{-1}$, and the maximum  intensity of 0.5.

The derived synthetic spectra for \civ\ troughs at specific line of sight y are shown as the histograms in the right panels of Figure \ref{fig:N_v}.
The synthetic absorption spectra we obtained shows a diversity in the trough widths. Examples of broad absorption troughs can be seen in panels (a) and (d) of Figure \ref{fig:N_v}, where the trough's width spans over more than $2000$ \kms. However, many of the troughs we obtain are narrower, 
an example of which can be seen in panel (c) of Figure \ref{fig:N_v}.
We note that the velocity width of observed quasar outflows, with a width larger than 1000 km/s,  has a strong peak around 1500 km/s (see section 4.3 and figure 6 in 
\citealt{trump2006}).
Finally, we also find marginally broad troughs of widths $\lsim2000$ \kms, an example of which is seen in panel (c) of Figure \ref{fig:N_v}. A more systematic analysis of the lines of sight in terms of BAL widths statistics as well as a direct comparison to observations will be presented in a follow-up paper (Xinfeng et al. in preparation).

\section{Conclusions}

In this work we constructed a numerical model for the interaction of AGN wind
with galactic clouds in an attempt to explain the observed BAL QSO spectra. 
The consideration of this model is motivated by recent observations that 
seem to indicate that a substantial fraction of the BAL outflows are located
at large distances from the central AGN - tens to hundreds parsecs.  

The wind-cloud interaction was modelled using 2D hydrodynamical simulations that incorporate
a simple cooling function.  The evolution of the system was followed for sufficiently long time to
allow complete destruction of the cloud by the engulfing wind and further acceleration of its debris to high velocities. 
The output data of the simulation was rearranged and integrated along lines of sight in the direction of
the wind propagation, and then used to compute synthetic spectra by accounting for absorption of
the AGN continuum radiation by the accelerated, cool cloud fragments.

We find that while a fraction of the cloud fragments reach velocities close to the injected wind velocity,
the cool dense cloudlets that can absorb the quasar light reach significantly lower 
velocities (as seen in Figs. \ref{fig:n-v} and \ref{fig:n-v_case2}), owing to the complex flow pattern 
within the Mach cone.  We also find that the mean number of absorbing cloudlets along lines of sight that
intersect the cloud debris depends on the density and radius of the pre-crashed cloud. 
For a wind velocity $v_w\simeq 0.1c$ and sufficiently dense cloud the synthetic spectra that we obtained 
mimic the features of observed BAL spectra in terms of acquired velocities (blue-shifted wavelengths) and in terms of column
densities (depths of the absorption troughs). In some of the lines of sight tested we find broad trough widths up to $\sim3000\textrm{km}\textrm{s}^{-1}$, whereas along other lines we find narrower troughs. Further analysis and direct comparisons to observations will be presented in 
a follow-up paper (Xinfeng et al. in preparation). 

M.Z.H. and A.L. acknowledge supported by the Israel Science Foundation Grant 1114/17. A.L. thanks Hagai Netzer for discussion and useful comments. M.Z.H. thanks Roy Gomel for the technical support. N.A. and X.X. acknowledge support from  NASA STScI grants GO 14777, 14242, 14054, and 14176, and NASA ADAP 48020.


\begin{thebibliography}{}
\makeatletter
\relax
\def\mn@urlcharsother{\let\do\@makeother \do\$\do\&\do\#\do\^\do\_\do\%\do\~}
\def\mn@doi{\begingroup\mn@urlcharsother \@ifnextchar [ {\mn@doi@}
  {\mn@doi@[]}}
\def\mn@doi@[#1]#2{\def\@tempa{#1}\ifx\@tempa\@empty \href
  {http://dx.doi.org/#2} {doi:#2}\else \href {http://dx.doi.org/#2} {#1}\fi
  \endgroup}
\def\mn@eprint#1#2{\mn@eprint@#1:#2::\@nil}
\def\mn@eprint@arXiv#1{\href {http://arxiv.org/abs/#1} {{\tt arXiv:#1}}}
\def\mn@eprint@dblp#1{\href {http://dblp.uni-trier.de/rec/bibtex/#1.xml}
  {dblp:#1}}
\def\mn@eprint@#1:#2:#3:#4\@nil{\def\@tempa {#1}\def\@tempb {#2}\def\@tempc
  {#3}\ifx \@tempc \@empty \let \@tempc \@tempb \let \@tempb \@tempa \fi \ifx
  \@tempb \@empty \def\@tempb {arXiv}\fi \@ifundefined
  {mn@eprint@\@tempb}{\@tempb:\@tempc}{\expandafter \expandafter \csname
  mn@eprint@\@tempb\endcsname \expandafter{\@tempc}}}

\bibitem[\protect\citeauthoryear{{Aoki}, {Oyabu}, {Dunn}, {Arav}, {Edmonds},
  {Korista}, {Matsuhara}  \& {Toba}}{{Aoki} et~al.}{2011}]{Aoki11}
{Aoki} K.,  {Oyabu} S.,  {Dunn} J.~P.,  {Arav} N.,  {Edmonds} D.,  {Korista}
  K.~T.,  {Matsuhara} H.,   {Toba} Y.,  2011, \pasj, \href
  {http://adsabs.harvard.edu/abs/2011PASJ...63S.457A} {63, 457}

\bibitem[\protect\citeauthoryear{{Arav}, {Liu}, {Xu}, {Stidham}, {Benn}  \&
  {Chamberlain}}{{Arav} et~al.}{2018}]{Arav18}
{Arav} N.,  {Liu} G.,  {Xu} X.,  {Stidham} J.,  {Benn} C.,   {Chamberlain} C.,
  2018, \mn@doi [\apj] {10.3847/1538-4357/aab494}, \href
  {http://adsabs.harvard.edu/abs/2018ApJ...857...60A} {857, 60}
  
  
\bibitem[\protect\citeauthoryear{Arav et al.}{1999}]{Arav99} Arav N., Becker R.~H., Laurent-Muehleisen S.~A., Gregg M.~D., White R.~L., Brotherton M.~S., de Kool M., 1999, ApJ, 524, 566
\bibitem[Arav et al.(2001)]{Arav01} Arav, N., de Kool, M., Korista, K.~T., et al.\ 2001, \apj, 561, 118  
\bibitem[Arav et al.(2013)]{Arav13} Arav, N., Borguet, B., Chamberlain, C., Edmonds, D., \& Danforth, C.\ 2013, \mnras, 436, 3286
\bibitem[\protect\citeauthoryear{Arav et al.}{2015}]{Arav15} Arav N., et al., 2015, A\&A, 577, A37

\bibitem[\protect\citeauthoryear{{Bautista}, {Dunn}, {Arav}, {Korista}, {Moe}
  \& {Benn}}{{Bautista} et~al.}{2010}]{Bautista10}
{Bautista} M.~A.,  {Dunn} J.~P.,  {Arav} N.,  {Korista} K.~T.,  {Moe} M.,
  {Benn} C.,  2010, \mn@doi [\apj] {10.1088/0004-637X/713/1/25}, \href
  {http://adsabs.harvard.edu/abs/2010ApJ...713...25B} {713, 25}

\bibitem[\protect\citeauthoryear{{Borguet}, {Arav}, {Edmonds}, {Chamberlain}
  \& {Benn}}{{Borguet} et~al.}{2013}]{Borguet13}
{Borguet} B.~C.~J.,  {Arav} N.,  {Edmonds} D.,  {Chamberlain} C.,   {Benn} C.,
  2013, \mn@doi [\apj] {10.1088/0004-637X/762/1/49}, \href
  {http://adsabs.harvard.edu/abs/2013ApJ...762...49B} {762, 49}

\bibitem[\protect\citeauthoryear{{Chamberlain} \& {Arav}}{{Chamberlain} \&
  {Arav}}{2015}]{Chamberlain15b}
{Chamberlain} C.,  {Arav} N.,  2015, \mn@doi [\mnras] {10.1093/mnras/stv1979},
  \href {http://adsabs.harvard.edu/abs/2015MNRAS.454..675C} {454, 675}

\bibitem[\protect\citeauthoryear{{Chamberlain}, {Arav}  \&
  {Benn}}{{Chamberlain} et~al.}{2015}]{Chamberlain15}
{Chamberlain} C.,  {Arav} N.,   {Benn} C.,  2015, \mn@doi [\mnras]
  {10.1093/mnras/stv572}, \href
  {http://adsabs.harvard.edu/abs/2015MNRAS.450.1085C} {450, 1085}

\bibitem[\protect\citeauthoryear{{Contopoulos}, {Kazanas}  \&
  {Fukumura}}{{Contopoulos} et~al.}{2017}]{Contopoulos17}
{Contopoulos} I.,  {Kazanas} D.,   {Fukumura} K.,  2017, \mn@doi [\mnras]
  {10.1093/mnrasl/slx123}, \href
  {http://adsabs.harvard.edu/abs/2017MNRAS.472L..20C} {472, L20}

\bibitem[\protect\citeauthoryear{{Dai}, {Shankar}  \& {Sivakoff}}{{Dai}
  et~al.}{2008}]{Dai08}
{Dai} X.,  {Shankar} F.,   {Sivakoff} G.~R.,  2008, \mn@doi [\apj]
  {10.1086/523688}, \href {http://adsabs.harvard.edu/abs/2008ApJ...672..108D}
  {672, 108}

\bibitem[\protect\citeauthoryear{{Dunn} et~al.,}{{Dunn} et~al.}{2010}]{Dunn10a}
{Dunn} J.~P.,  et~al., 2010, \mn@doi [\apj] {10.1088/0004-637X/709/2/611},
  \href {http://adsabs.harvard.edu/abs/2010ApJ...709..611D} {709, 611}

\bibitem[\protect\citeauthoryear{{Faucher-Gigu{\`e}re}, {Quataert}  \&
  {Murray}}{{Faucher-Gigu{\`e}re} et~al.}{2012}]{Faucher-Giguere12}
{Faucher-Gigu{\`e}re} C.-A.,  {Quataert} E.,   {Murray} N.,  2012, \mn@doi
  [\mnras] {10.1111/j.1365-2966.2011.20120.x}, \href
  {http://adsabs.harvard.edu/abs/2012MNRAS.420.1347F} {420, 1347}

\bibitem[Ferland et al.(2017)]{Ferland17} Ferland, G.~J., Chatzikos, M., Guzmn, F., et al.\ 2017, RMxAA, 53, 385

\bibitem[Grevesse et al.(2010)]{GASS10} Grevesse, N., Asplund, M., Sauval, A.~J., \& Scott, P.\ 2010, \apss, 328, 179 

\bibitem[\protect\citeauthoryear{{Gnat} \& {Ferland}}{{Gnat} \&
  {Ferland}}{2012}]{gnat2012}
{Gnat} O.,  {Ferland} G.~J.,  2012, \mn@doi [\apjs]
  {10.1088/0067-0049/199/1/20}, \href
  {https://ui.adsabs.harvard.edu/abs/2012ApJS..199...20G} {199, 20}

\bibitem[\protect\citeauthoryear{{Hewett} \& {Foltz}}{{Hewett} \&
  {Foltz}}{2003}]{Hewett03}
{Hewett} P.~C.,  {Foltz} C.~B.,  2003, \mn@doi [\aj] {10.1086/368392}, \href
  {http://adsabs.harvard.edu/abs/2003AJ....125.1784H} {125, 1784}

\bibitem[\protect\citeauthoryear{{Mignone}, {Bodo}, {Massaglia}, {Matsakos},
  {Tesileanu}, {Zanni}  \& {Ferrari}}{{Mignone} et~al.}{2007}]{mignone2007}
{Mignone} A.,  {Bodo} G.,  {Massaglia} S.,  {Matsakos} T.,  {Tesileanu} O.,
  {Zanni} C.,   {Ferrari} A.,  2007, \mn@doi [\apjs] {10.1086/513316}, \href
  {http://adsabs.harvard.edu/abs/2007ApJS..170..228M} {170, 228}

\bibitem[\protect\citeauthoryear{{Miller}, {Arav}, {Xu}, {Kriss}, {Plesha},
  {Benn}  \& {Liu}}{{Miller} et~al.}{2018}]{Miller18}
{Miller} T.~R.,  {Arav} N.,  {Xu} X.,  {Kriss} G.~A.,  {Plesha} R.~J.,  {Benn}
  C.,   {Liu} G.,  2018, \mn@doi [\apj] {10.3847/1538-4357/aad817}, \href
  {https://ui.adsabs.harvard.edu/#abs/2018ApJ...865...90M} {865, 90}

\bibitem[\protect\citeauthoryear{{Moe}, {Arav}, {Bautista}  \& {Korista}}{{Moe}
  et~al.}{2009}]{Moe09}
{Moe} M.,  {Arav} N.,  {Bautista} M.~A.,   {Korista} K.~T.,  2009, \mn@doi
  [\apj] {10.1088/0004-637X/706/1/525}, \href
  {http://adsabs.harvard.edu/abs/2009ApJ...706..525M} {706, 525}

\bibitem[\protect\citeauthoryear{{Murray}, {Chiang}, {Grossman}  \&
  {Voit}}{{Murray} et~al.}{1995}]{Murray95}
{Murray} N.,  {Chiang} J.,  {Grossman} S.~A.,   {Voit} G.~M.,  1995, \mn@doi
  [\apj] {10.1086/176238}, \href
  {http://adsabs.harvard.edu/abs/1995ApJ...451..498M} {451, 498}

\bibitem[\protect\citeauthoryear{{Ostriker}, {Choi}, {Ciotti}, {Novak}  \&
  {Proga}}{{Ostriker} et~al.}{2010}]{Ostriker10}
{Ostriker} J.~P.,  {Choi} E.,  {Ciotti} L.,  {Novak} G.~S.,   {Proga} D.,
  2010, \mn@doi [\apj] {10.1088/0004-637X/722/1/642}, \href
  {http://adsabs.harvard.edu/abs/2010ApJ...722..642O} {722, 642}

\bibitem[\protect\citeauthoryear{{Proga}, {Stone}  \& {Kallman}}{{Proga}
  et~al.}{2000}]{Proga00}
{Proga} D.,  {Stone} J.~M.,   {Kallman} T.~R.,  2000, \mn@doi [\apj]
  {10.1086/317154}, \href {http://adsabs.harvard.edu/abs/2000ApJ...543..686P}
  {543, 686}
  
  \bibitem[Savage \& Sembach(1991)]{Savage91} Savage, B.~D., \& Sembach, K.~R.\ 1991, \apj, 379, 245 

\bibitem[\protect\citeauthoryear{{Sutherland} \& {Dopita}}{{Sutherland} \&
  {Dopita}}{1993}]{sutherland1993}
{Sutherland} R.~S.,  {Dopita} M.~A.,  1993, \mn@doi [\apjs] {10.1086/191823},
  \href {https://ui.adsabs.harvard.edu/abs/1993ApJS...88..253S} {88, 253}

\bibitem[\protect\citeauthoryear{{Thompson}, {Fabian}, {Quataert}  \&
  {Murray}}{{Thompson} et~al.}{2015}]{Thompson15}
{Thompson} T.~A.,  {Fabian} A.~C.,  {Quataert} E.,   {Murray} N.,  2015,
  \mn@doi [\mnras] {10.1093/mnras/stv246}, \href
  {https://ui.adsabs.harvard.edu/abs/2015MNRAS.449..147T} {449, 147}
  
  \bibitem[Trump et al.(2006)]{trump2006} Trump, J.~R., Hall, P.~B., Reichard, T.~A., et al.\ 2006, \apjs, 165, 1
  
  \bibitem[Weymann et al.(1991)]{Weymann91} Weymann, R.~J., Morris, S.~L., Foltz, C.~B., \& Hewett, P.~C.\ 1991, \apj, 373, 23
  \mn@doi [\apj] {10.1086/170020}, \href
  {http://articles.adsabs.harvard.edu/abs/1991ApJ...373...23W} {373, 23}

\bibitem[\protect\citeauthoryear{{Xu}, {Arav}, {Miller}  \& {Benn}}{{Xu}
  et~al.}{2018a}]{Xu18b}
{Xu} X.,  {Arav} N.,  {Miller} T.,   {Benn} C.,  2018a, preprint, \href
  {https://ui.adsabs.harvard.edu/#abs/2018arXiv180501544X} {p.
  arXiv:1805.01544} (\mn@eprint {arXiv} {1805.01544})

\bibitem[\protect\citeauthoryear{{Xu}, {Arav}, {Miller}  \& {Benn}}{{Xu}
  et~al.}{2018b}]{Xu18a}
{Xu} X.,  {Arav} N.,  {Miller} T.,   {Benn} C.,  2018b, \mn@doi [\apj]
  {10.3847/1538-4357/aab7ea}, \href
  {https://ui.adsabs.harvard.edu/#abs/2018ApJ...858...39X} {858, 39}

\bibitem[\protect\citeauthoryear{{Yuan}, {Ostriker}, {Yoon}, {Li}, {Ciotti},
  {Gan}, {Ho}  \& {Guo}}{{Yuan} et~al.}{2018}]{Yuan18}
{Yuan} F.,  {Ostriker} J.~P.,  {Yoon} D.,  {Li} Y.-P.,  {Ciotti} L.,  {Gan}
  Z.-M.,  {Ho} L.~C.,   {Guo} F.,  2018, arXiv e-prints, \href
  {https://ui.adsabs.harvard.edu/abs/2018arXiv180705488Y} {p. arXiv:1807.05488}

\bibitem[\protect\citeauthoryear{{Zeilig-Hess}, {Levinson}  \&
  {Nakar}}{{Zeilig-Hess} et~al.}{2019}]{zeilig2019}
{Zeilig-Hess} M.,  {Levinson} A.,   {Nakar} E.,  2019, \mn@doi [\mnras]
  {10.1093/mnras/sty3034}, \href
  {https://ui.adsabs.harvard.edu/abs/2019MNRAS.482.4642Z} {482, 4642}

\bibitem[\protect\citeauthoryear{{Zubovas} \& {Nayakshin}}{{Zubovas} \&
  {Nayakshin}}{2014}]{Zubovas14}
{Zubovas} K.,  {Nayakshin} S.,  2014, \mn@doi [\mnras] {10.1093/mnras/stu431},
  \href {http://adsabs.harvard.edu/abs/2014MNRAS.440.2625Z} {440, 2625}

\makeatother
\end{thebibliography}

\end{document}